%% file: usenix.tex
\documentclass[letterpaper,twocolumn,10pt]{article}
\usepackage{usenix-2020-09}

\usepackage{tikz}
\usepackage{amsmath}

\usepackage{booktabs}  
\usepackage{hyperref}
\usepackage{breakurl}
\usepackage{comment}
\usepackage{enumitem}
\usepackage[ruled,vlined,linesnumbered]{algorithm2e}
\usepackage{subfigure}
\usepackage{tikz}
\usepackage{pifont}
\usepackage{bbding}
\usepackage{multirow}
\usepackage{makecell}
\usepackage{authblk}

\usepackage{listings}
\usepackage{xcolor}
\definecolor{dkgreen}{rgb}{0,0.6,0}
\definecolor{gray}{rgb}{0.5,0.5,0.5}
\definecolor{mauve}{rgb}{0.58,0,0.82}

\definecolor{codegreen}{rgb}{0,0.6,0}
\definecolor{codegray}{rgb}{0.5,0.5,0.5}
\definecolor{codepurple}{rgb}{0.58,0,0.82}
\definecolor{backcolour}{rgb}{0.95,0.95,0.92}

\begin{document}

\date{}

\title{\Large \bf ONCache: A Cache-Based Low-Overhead Container Overlay Network}

\setlength{\affilsep}{0.1em}
\makeatletter
\renewcommand\AB@affilsepx{, \protect\Affilfont}
\renewcommand\Affilfont{\itshape}
\makeatother
\author[1]{Shengkai Lin} 
\author[1]{Shizhen Zhao}
\author[1]{Peirui Cao}
\author[1]{Xinchi Han}
\author[2]{\\ Quan Tian}
\author[2]{Wenfeng Liu}
\author[2]{Qi Wu}
\author[2]{Donghai Han}
\author[1]{Xinbing Wang}

\affil[1]{Shanghai Jiao Tong University}
\affil[2]{VMware}
\maketitle

\input{sections/0Abstract.tex}

\input{sections/1Introduction.tex}

\input{sections/2Background.tex}

\input{sections/3Overview.tex}

\input{sections/4Design.tex}

\input{sections/6Evaluation.tex}

\input{sections/7Discussion.tex}

\input{sections/8Relatedworks.tex}
\input{sections/9Conclusion.tex}

\bibliographystyle{plain}
\bibliography{bibliography}

\input{sections/10appendix}

\end{document}

%% file: sections/0Abstract.tex
\begin{abstract}
Recent years have witnessed a widespread adoption of containers. While containers simplify and accelerate application development, existing container network technologies either incur significant overhead, which hurts performance for distributed applications, or lose flexibility or compatibility, which hinders the widespread deployment in production.

We carefully analyze the kernel data path of an overlay network, quantifying the time consumed by each segment of the data path and identifying the \emph{extra overhead} in an overlay network compared to bare metal. We observe that this extra overhead generates repetitive results among packets, which inspires us to introduce caches within an overlay network.

We design and implement ONCache (\textbf{O}verlay \textbf{N}etwork \textbf{Cache}), a cache-based container overlay network, to eliminate the extra overhead while maintaining flexibility and compatibility. 
We implement ONCache using the extended Berkeley Packet Filter (eBPF) with only 524 lines of code, and integrate it as a plugin of Antrea.
With ONCache, containers attain networking performance akin to that of bare metal.
Compared to the standard overlay networks, ONCache improves throughput and request-response transaction rate by 12\% and 36\% for TCP (20\% and 34\% for UDP), respectively, while significantly reducing per-packet CPU overhead. 
Popular distributed applications also benefit from ONCache. 

\end{abstract}

%% file: sections/1Introduction.tex
\section{Introduction}
Containers~\cite{docker, cpe.5668} are becoming increasingly popular for distributed application deployment due to their flexibility and lightweight nature. By bundling an application along with all its dependencies and configuration files into one container image, containers bring forth the ability to \emph{build once and run anywhere}. Furthermore, containers leverage the host OS kernel instead of emulating an entire OS, and thus are more lightweight than virtual machines. Container orchestration tools such as Kubernetes~\cite{k8s}, Docker Swarm~\cite{swarm}, etc., further reduce container management complexity by facilitating features like auto-deployment, auto-scaling, and auto-high-availability~\cite{Casalicchio2019}.

Container networks play a key role in enabling communication between containers and supporting distributed applications. Ideally, a container network should offer low-overhead, high-performance networking while preserving flexibility and compatibility. However, none of the existing container networks fully satisfy these expectations, as outlined in Table~\ref{tab:compare}.

\begin{table}[t]\footnotesize
	\centering
    \begin{tabular}{l|c|c|c}
    \toprule
\textbf{Technology} & \textbf{Performance} & \textbf{Flexibility} & \textbf{Compatibility} \\ \midrule
Host              & \Checkmark                    & \XSolidBrush                    & \Checkmark                     \\
Bridge            & \Checkmark                    & \XSolidBrush                    & \Checkmark                     \\
Macvlan           & \Checkmark                    & \XSolidBrush                    & \Checkmark                     \\
IPvlan           & \Checkmark                    & \XSolidBrush                    & \Checkmark                     \\
SR-IOV~\cite{sriov}           & \Checkmark                    & \XSolidBrush                    & \Checkmark                     \\
Overlay           & \XSolidBrush                    & \Checkmark                    & \Checkmark                     \\
Falcon~\cite{osti_10297232} & \XSolidBrush                    & \Checkmark                    & \Checkmark                     \\
Slim~\cite{zhuo2019slim}              & \Checkmark                    & \Checkmark                    & \XSolidBrush                     \\
\textbf{ONCache} & \Checkmark                    & \Checkmark                    & \Checkmark                    \\
\bottomrule
    \end{tabular}
    \setlength{\belowcaptionskip}{-10pt}
\caption{\textit{Compare container networking technologies.}}
\label{tab:compare}
\end{table}

\textbf{Hurt flexibility.} Containers in the host network share the host IP address and must coordinate ports on the host, constraining deployment flexibility. Bridge networks or device virtualization techniques (such as Macvlan/IPvlan/SR-IOV~\cite{sriov}) allow containers to have their own IP addresses and directly forward container packets to the underlay network. However, these networks require either all hosts to be in the same L2 network or the underlay network to be capable of routing container packets, imposing restrictions on container placement, migration, or network configuration.

\textbf{Poor performance and high overhead.} Container overlay networks utilizing tunneling techniques (e.g., VXLAN~\cite{rfc7348}) completely decouple containers and the underlay network, facilitating the configuration of container IP addresses regardless of the underlay network. Nonetheless, it incurs significant overhead~\cite{8485865, zhuo2019slim, osti_10297232, lei2019tackling}. Our experiments show that compared to bare metal, the throughput of a single TCP flow in a tunnel-based overlay network is about 11\% lower, with the TCP request-response transaction rate being around 29\% lower. Moreover, CPU utilization is much higher in both experiments. While certain techniques~\cite{rfs, osti_10297232, 10177405} endeavor to enhance overlay networks by distributing ingress packet processing across multiple CPU cores, the processing overhead of overlay networks remains significant.

\textbf{Hurt compatibility.} Slim~\cite{zhuo2019slim} and its follow-up works~\cite{9860379, 10198904, 10143220} implement overlay networks via socket replacement, enabling containerized applications to utilize the host's sockets. While they maintain low overhead and some degree of flexibility in overlay networks, they do not inherently support non-connection-based protocols (e.g., UDP, ICMP) or container live migration. Additionally, they may conflict with tunneling-headers-based policies in the underlay network.

To attain high performance, flexibility, and compatibility at the same time, we design and implement ONCache (\textbf{O}verlay \textbf{N}etwork \textbf{Cache}), a container overlay network featuring a cache-based and low-overhead fast path. ONCache can be seamlessly integrated into existing standard overlay networks and has been tested with Antrea~\cite{antrea} and Flannel~\cite{flannel}. 

Our detailed contributions are summarized as follows:

\begin{enumerate}
[itemsep=2pt,topsep=0pt,parsep=0pt]
\item 
We carefully analyze the kernel data path of popular container overlay networks (Antrea and Cilium) alongside bare metal, quantifying the time consumption of each segment of the overhead. Our analysis reveals the \emph{extra overhead} incurred by container overlay networks. 
However, the overhead disperses across the kernel data path, presenting a significant challenge in mitigating it while preserving essential functionalities.
(\S \ref{sec:overheadana})

\item 
We identify the \emph{invariance property} within the extra overhead, indicating that the processing results of the extra overhead are repetitive among packets. This property inspires us to use caches in overlay networks. (\S \ref{sec:motivation})

\item 
We propose a cache-based fast path for container overlay networks. The fast path bypasses the extra overhead without losing flexibility and compatibility. (\S \ref{sec:design})

\item We evaluate ONCache with microbenchmarks and popular distributed applications. Results demonstrate significant improvement of ONCache in throughput (12\% higher with iperf3), latency (36\% higher transaction rate), and CPU utilization (26\% lower per-transaction) compared to the standard overlay networks, approaching performance akin to that of bare metal. Applications such as Memcached, PostgreSQL, and Nginx all derive benefits from ONCache. (\S \ref{sec:evaluation}) 
\end{enumerate}

ONCache’s source code is publicly available at \url{https://github.com/nothepeople/ONCache}.

\emph{This work does not raise any ethical issues.}

%% file: sections/2Background.tex
\begin{table*}[t]\footnotesize
    \centering
    \begin{tabular}{l|l|rrrr|l|rrrr}
    \toprule
     & \multicolumn{5}{c|}{\textbf{Egress}} & \multicolumn{5}{c}{\textbf{Ingress}} \\
    \midrule
    \textbf{Data path} & Overhead type & Antrea & Cilium & BM & \textbf{Ours} & Overhead type & Antrea & Cilium & BM & \textbf{Ours} \\
    \hline
    \multirow{4}{*}{\textbf{\begin{tabular}[c]{@{}l@{}}Application\\ network stack\end{tabular}}} & skb allocation & 1505 & 1566 & 1461 & \textbf{1509} & skb releasing & 715 & 818 & 780 & \textbf{714} \\
     & Conntrack & 778 & 0 & 788 & \textbf{763} & Conntrack & 616 & 0 & 600 & \textbf{592} \\
     & Netfilter & 0 & 0 & 305 & \textbf{0} & Netfilter & 0 & 0 & 173 & \textbf{0} \\
     & Others & 423 & 560 & 547 & \textbf{519} & Others & 838 & 1016 & 979 & \textbf{982} \\
    \hline
    \textbf{Veth pair*} & NS traversing & 562 & 594 & & \textbf{489} & NS traversing & 400 & & & \\
    \hline
    \textbf{eBPF*} & eBPF & & 1513 & & \textbf{511} & eBPF & & 1429 & & \textbf{289} \\
    \hline
    \multirow{3}{*}{\textbf{Open vSwitch*}} & Conntrack & 872 & & & & Conntrack & 758 & & & \\
     & Flow matching & 354 & & & & Flow matching & 308 & & & \\
     & Action execution & 92 & & & & Action execution & 66 & & & \\
    \hline
     \multirow{4}{*}{\textbf{\begin{tabular}[c]{@{}l@{}}VXLAN\\ network stack*\end{tabular}}} & Conntrack & 0 & 471 & & & Conntrack & 0 & 271 & & \\
     & Netfilter & 667 & 421 & & & Netfilter & 466 & 303 & & \\
     & Routing & 50 & 468 & & & Routing & 294 & 554 & & \\
     & Others & 319 & 127 & & & Others & 619 & 444 & & \\
    \hline
    \textbf{Link layer} & Link layer & 1858 & 1763 & 1799 & \textbf{1700} & Link layer & 2790 & 2848 & 2800 & \textbf{2737} \\
    \hline
    \textbf{Sum} & & 7479 & 7483 & 4900 & \textbf{5491} & & 7869 & 7683 & 5332 & \textbf{5315} \\
    \midrule
    \textbf{Latency} ($\mu$s) & & 22.97 & 23.15 & 16.57 & \textbf{17.49} & & 22.97 & 23.15 & 16.57 & \textbf{17.49} \\
    \bottomrule
    \end{tabular}
    \setlength{\abovecaptionskip}{5pt}
    \setlength{\belowcaptionskip}{-8pt}
    \caption{\textit{Overhead breakdown of different networks. All values, except for the end-to-end latency in the last row, are in nanoseconds. BM stands for bare metal. Ours, i.e., ONCache, is proposed in \S\ref{sec:design} and evaluated in \S\ref{sec:evaluation}. ``*'' denotes the extra overhead compared to bare metal. Due to the limitations of the measurement tool, there is an error of about 200 ns.}}
    \label{tab:overhead}
\end{table*}

\section{Background and Motivation}
\label{sec:background}
We analyze inter-host container networks, where overlay networks stand out for the superior flexibility and compatibility. After quantifying the detailed overhead of overlay networks and exploring recent efforts aimed at mitigating this overhead, we provide the motivation of ONCache.

\subsection{Inter-host Container Networks}
Container networks play a key role in facilitating communication among containers and with the external world. Below, we focus on inter-host container networks, which are classified based on the IP addresses utilized by \emph{applications} and the \emph{physical network}.

\textbf{Both use host IP addresses.} Host networks fall into this category. 
Containers within host networks share the host's network namespace and utilize the host's IP address. 
Host networks offer performance comparable to that of bare metal (where applications directly execute on the host)~\cite{8485865}.
Nonetheless, host networks demand coordination of ports among containers residing on the same host, thereby substantially limiting application flexibility.
This limitation renders host networks seldom used in production~\cite{k8snetwork}.

\textbf{Both use container IP addresses.} Bridge networks and device virtualization techniques, such as Macvlan, IPvlan, SR-IOV, fall into this category. In these networks, each container is assigned a unique IP address and is isolated from the host namespace. 
In bridge networks, all the containers are connected to Open vSwitch (OVS)~\cite{antrea}, bridge~\cite{flannel}, or similar entity, relying on them to forward packets to the physical network. Networks employing device virtualization techniques emulate multiple virtual interfaces with different IP addresses on each physical interface, with each virtual interface being attached to a container. These networks deliver performance akin to that of bare metal due to their simple data paths.

However, owing to their utilization of container IP addresses in the physical network, the above approaches exhibit poor flexibility. To route container IP addresses, one must either (1) place all hosts into one L2 network or (2) install and manage routing rules for container IP addresses within the physical network.
These requirements impose constraints on container placement, migration or network configuration. 

\textbf{App uses container IP addresses and the physical network uses host IP addresses.} Overlay networks fall within this category. 
In overlay networks, each container has its own IP address, and container packets are encapsulated using a tunneling protocol (e.g., VXLAN~\cite{rfc7348} or GENEVE~\cite{rfc8926}) before being transmitted to the physical network. Overlay networks completely decouple containers from the physical network, thereby ensuring flexibility for both. However, tunneling incurs extra overhead and hurts performance (see \S \ref{sec:overheadana}).

\textbf{\emph{Takeaway} \#1} Overlay networks decouple applications from the physical network, allowing for application deployment without concern for physical network configuration, and vice versa.
Given the superior flexibility of overlay networks, we focus on analyzing and improving overlay networks.

\subsection{Analysis of Overlay Network Overhead}
\label{sec:overheadana}
Prior studies~\cite{8485865, zhuo2019slim, osti_10297232, lei2019tackling} have shown the poor performance of overlay networks. We further quantify the overhead of two common overlay networks: Antrea~\cite{antrea} (v1.9.0) and Cilium~\cite{cilium} (v1.12.4), in comparison with bare metal.

Our analysis of the networks follows these steps.
First, we qualitatively deconstruct the egress and ingress data paths using flame graphs~\cite{flamegraph}, which provide comprehensive function call stacks on the data path. 
Next, we identify the exact functions (alongside their call stacks) that represent each segment of the overhead from the flame graphs. 
Finally, we measure the execution time for each function using the extended Berkeley Packet Filter (eBPF) and subsequently calculate the execution time for each segment of the overhead.
The detailed analysis method is provided in Appendix~\ref{app:bcctiming}.

Table~\ref{tab:overhead} presents the deconstructed overhead and the execution time for each segment of the overhead. Each listed execution time is the average of all the timing samples within a 1-second 1-byte TCP request-response (RR) test. 
We primarily analyze the overhead within bare metal and Antrea (``*'' denotes the extra overhead compared to bare metal), while the distinct overhead within Cilium is also discussed.
ONCache (ours) is analyzed in \S\ref{sec:evaluation}.

\textbf{Application network stack.} 
The application network stack directly interacts with applications. On the egress path, the application writes data to the socket. The application network stack then allocates a socket buffer (i.e., skb) for the data and encapsulates the packet layer by layer. Conversely, on the ingress path, the network stack decapsulates the headers and releases the socket buffer. Additionally, netfilter~\cite{netfilter} filters packets on both the ingress and egress paths. Conntrack~\cite{conntrack}, as a component of netfilter, tracks connection (not TCP connections, defined by, for example, 5-tuple: source IP address, source port, destination IP address, destination port, and transport protocol) states and facilitates stateful filters to address probes and denial-of-service attacks.
The execution of netfilter and conntrack depends on the system configuration.

\textbf{Veth pair*.} 
Containers run applications within individual network namespaces and employ veth pairs~\cite{veth} to connect with the host namespace. Traversing veth pairs incurs overhead, which includes transmit queuing on the sender veth and software interrupt scheduling on the receiver veth. In Cilium, although the overhead of traversing veth pairs can be mitigated on the ingress path through bpf (Berkeley Packet Filter) redirect~\cite{redirectpeer}, it still exists on the egress path~\cite{ebpfhostrouting}.

\textbf{OVS*.} 
OVS filters packets and intra-host routes packets between containers and the host interface. These filtering and routing actions are described by flows. The overhead in OVS can be categorized into connection tracking, flow matching, and action execution.
Despite OVS employing a cache to expedite flow matching~\cite{15-ovs}, connection tracking still consumes a substantial amount of CPU time.
Cilium implements similar functionalities as OVS in eBPF~\cite{ciliumebpf}. However, the overall eBPF execution time is similar to that of OVS.

\textbf{VXLAN\footnote{We take VXLAN as an example in following discussion. The analysis is similar for other tunneling protocols.} network stack*.} 
The VXLAN network stack performs egress and ingress routing for VXLAN packets and encapsulates or decapsulates the outer headers (including VXLAN header, UDP header, IP header, and MAC header) to or from the packets.
In Antrea, VXLAN routing is accelerated by OVS, resulting in notably low routing overhead.
Additionally, conntrack tracks connections, and netfilter filters packets at this stage.

\textbf{Link layer.}
On the egress path, the link layer queues and transmits packets
to the network; on the ingress path, it allocates socket buffers and receives packets from the network.

\textbf{\emph{Takeaway} \#2} The \emph{extra overhead} in overlay networks, including veth pair, OVS (or similar entity), and the VXLAN network stack, incurs significant overhead and degrades network performance.
Given its dispersion across various parts of the data path, mitigating the overhead while preserving essential functionalities presents a significant challenge.

\subsection{Related Works on Overlay Networks}
\label{sec:relatedworks}
There are some recent efforts that aim at improving the performance of overlay networks.

\textbf{Socket replacement.} Slim~\cite{zhuo2019slim} introduces a socket replacement mechanism that improves container overlay network performance. It establishes TCP connections in the host namespace and replaces TCP sockets in containers by those in the host namespace.
While improving TCP performance, Slim exhibits notable limitations. Firstly, it does not support non-connection-based protocols such as UDP and ICMP, or container live migration, thereby constraining its compatibility. Secondly, Slim packets are not tunneling packets, which impacts underlay network policies that match on tunneling headers. Furthermore, Slim must establish an overlay connection for service discovery during connection setup, resulting in prolonged setup time.
Some subsequent works~\cite{9860379, 10198904, 10143220} follow Slim's idea and improve connection setup time. However, they still suffer from the former two shortcomings. 

\textbf{CPU load balancing.} Linux scaling techniques, such as Receive Packet Steering (RPS) and Receive Flow Steering (RFS) (and their hardware counterparts Receive Side Scaling (RSS) and Accelerated RFS (aRFS))~\cite{rfs}, improve performance by distributing ingress packet processing across multiple CPU cores and trying to reduce cache miss rate. 
Falcon~\cite{osti_10297232} and mFlow~\cite{10177405} explore different strategies for distributing packet processing across multiple cores.
However, they improve network performance at the cost of higher CPU utilization.
Additionally, they only take effects on the ingress path and cannot further improve performance if the bottleneck lies in the egress path.

\textbf{\emph{Takeaway} \#3} Despite many efforts to improve overlay networks, none of them simultaneously maintains low overhead, high performance, high flexibility and high compatibility. 

\subsection{Motivation: the Invariance Property}
\label{sec:motivation}
Based on the analysis in \S\ref{sec:overheadana}, we further classify the processing of the extra overhead into following types: connection tracking, packet filtering, intra-host routing, namespaces traversing and outer-header processing~\footnote{If Kubernetes ClusterIP Service~\cite{service} is enabled, inter-host container traffic also needs Network Address Translation (NAT), as discussed in \S\ref{sec:compatibilityconsiderations}.}. 
Among these overhead types, namespaces traversing does not yield any processing result, while the other four exhibit an \emph{invariance property}.

\textbf{Invariance in connection tracking.} 
To the best of our knowledge, all common connection trackers (e.g., those in netfilter~\cite{ayuso2006netfilter}, OVS~\cite{ovsconntrack}, Cilium~\cite{ciliumconntrack}) have an established state, which signifies that the connection tracker has observed two-way communication of a connection. Once in the established state, the connection does not switch to another state until its completion.

\textbf{Invariance in packet filtering.} 
Packet filters can be classified into two types: stateless and stateful. Stateless filters only match on packet fields, typically the 5-tuple. Stateful filters also take into account the connection state, which is provided by the connection tracker.
While OVS and netfilter contain numerous filters, the final decision to allow or deny packets with the same 5-tuple remains unchanged once the flow enters the established state.

\textbf{Invariance in intra-host routing.} 
If the container IP addresses are within the same subnet, layer-2 routing is enough, and the packet can be directly forwarded to the target interface. Otherwise, layer-3 routing is necessary and the MAC addresses should be modified before forwarding the packet.
In both scenarios, the intra-host routing decisions are invariant for packets destined to the same container IP address.

\textbf{Invariance in outer-header processing.} 
On the ingress path, the outer headers are simply removed after checking fields such as destination and Time-to-Live (TTL). The correctness of the payload is ensured by the checksum of the inner headers.
The egress path encapsulates the outer headers, whose most fields are related to routing and network configuration. On the same host, these fields are invariant for packets destined to the same container IP address:
(1) The outer MAC header stores the MAC addresses, VLAN, and Ethernet type, which are invariant.
(2) The IP addresses, protocol, TTL, differentiated services code point (DSCP), etc. in the outer IP header are invariant. Length, identification (ID), checksum fields vary among packets, but they can be easily updated with little computation.
(3) In the outer UDP header, the destination port (VXLAN set to 4789), and the checksum (VXLAN set to 0)\footnote{While UDP checksum is necessary for certain tunneling protocols such as GENEVE, the computational overhead is low.} are invariant~\cite{rfc7348}. 
The source port is calculated by simply hashing the 5-tuple of the inner headers. The length field can also be easily updated.
(4) The VXLAN header stores the VXLAN network identifier (VNI), which does not change in an overlay network.

\textbf{\emph{Takeaway} \#4} 
The extra overhead of overlay networks exhibits an invariance property.
This observation motivates us to introduce \emph{caches} to eliminate the extra overhead.

%% file: sections/3Overview.tex
\section{ONCache}
\label{sec:design}

ONCache is a container overlay network that offers a cache-based fast path with low overhead, high performance, flexibility, and compatibility. 
It is essentially a plug-in of standard overlay networks such as Antrea, Flannel. ONCache adopts a fail-safe design: when the fast path is unavailable, the traffic seamlessly falls back to the standard overlay network. 

\textbf{Define cache:} Inspired by the invariance property discussed in \S\ref{sec:motivation}, 
ONCache defines three caches—the egress cache, the ingress cache, and the filter cache—on each host to store the invariant results of the extra overhead. The detailed design of these caches is introduced in \S\ref{sec:cache}.

\textbf{Initialize cache:} ONCache relies on a standard overlay network to initialize the cache. In the event of a cache miss, ONCache passes the packet to the standard overlay network for processing and stores the processing results in caches if the packet's flow reaches the established state.
See \S \ref{sec:cachemiss} for the detailed design. 

\textbf{Utilize cache:} Upon a cache hit, ONCache filters the packet using the cached filtering results, constructs a tunneling packet using the cached outer headers, and then redirects the packet to the target interface by the cached routing decision. 
The cache-based fast path eliminates the \emph{expensive} repetitive processing of \emph{multiple layers}, as analyzed in \S\ref{sec:motivation}, by several lightweight cache lookups. 
The detailed design is described in \S\ref{sec:cachetrans}.

\textbf{Maintain cache:} ONCache ensures that the outdated cache entries are appropriately 
evicted when network-changing events (including container deletion, migration, and filter modification) happen. This is discussed in \S\ref{sec:cachecoherency}.

\textbf{Cache compatibility:} ONCache focuses on inter-host container networking and is also compatible with various traffic and features in the network. This is discussed in \S \ref{sec:compatibilityconsiderations}.

\textbf{Optional improvements:} We provide two optional improvements to further enhance  performance. As they require kernel or tunneling protocol modifications, they are not applied as default designs. This is described in \S\ref{sec:optimp}.

\begin{figure}[t]
    \centering
    \includegraphics[width=0.38\textwidth]{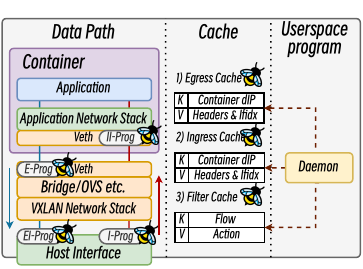}
    \setlength{\abovecaptionskip}{-5pt}
    \setlength{\belowcaptionskip}{-5pt}
    \caption{\textit{Architecture of ONCache. ONCache consists of 4 eBPF programs in the data path, 3 eBPF maps, and 1 user space program. The components with the eBPF logo (``bee'') represent eBPF programs or eBPF maps. dIP is short for destination IP address.
    }}
    \label{fig: architecture}

\end{figure}

\begin{table}[t]\footnotesize
	\centering
    \begin{tabular}{c|c|c}
    \toprule
\textbf{eBPF Program}  & \textbf{Abbr.} & \textbf{Hook Point} \\ \midrule
Egress-Prog & E-Prog &  TC ingress of the veth (host-side) \\
Ingress-Prog & I-Prog & TC ingress of the host interface\\
Egress-Init-Prog & EI-Prog & TC egress of the host interface \\
Ingress-Init-Prog & II-Prog & TC ingress of the veth (container-side) \\
\bottomrule
    \end{tabular}
\setlength{\abovecaptionskip}{-2pt}
\setlength{\belowcaptionskip}{-5pt}
\caption{\textit{Hook points of the four eBPF programs.}}
\label{tab:hook}
\vspace{-10pt}
\end{table}

\begin{figure*}[t]
\centering
    \begin{minipage}[b]{0.61\textwidth}
        \centering
        \includegraphics[width=1\textwidth]{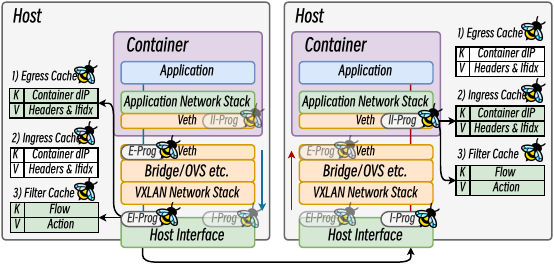}
        \setlength{\abovecaptionskip}{-0.6cm}
        \setlength{\belowcaptionskip}{-15pt}
        \makeatletter\def\@captype{figure}\makeatother\caption{\textit{The cache initialization process of ONCache. EI-Prog/II-Prog update the entries (green shaded) when the initialization requirements are met.
        }}
        \label{fig:init}
    \end{minipage}
    \hspace{0.4cm}
    \begin{minipage}[b]{0.27\textwidth}\small
        \centering
        \includegraphics[width=1\textwidth]{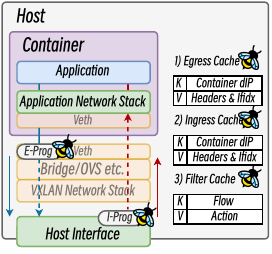}
        \setlength{\abovecaptionskip}{-0.6cm}
        \setlength{\belowcaptionskip}{-15pt}
        \makeatletter\def\@captype{figure}\makeatother\caption{\textit{Journey of an overlay packet in ONCache. EI-Prog/II-Prog are skiped over. The dash lines denotes the redirect path.}}
        \label{fig:packet_journey}
    \end{minipage}
\end{figure*}

For ease of implementation and deployment, we implement ONCache with eBPF \cite{ebpfio, 10.1145/3371038}. eBPF is a technology that runs sandboxed programs within a privileged context like the Linux kernel, thereby safely and efficiently extending the capabilities of the kernel.
eBPF programs are attached to kernel hook points (e.g., TC (traffic control) or XDP (eXpress Data Path)~\cite{hoiland2018express}) and are triggered each time the kernel executes through the hook points.  
eBPF stores data in a map data structure residing in kernel space, called eBPF map. Numerous prior works leverage eBPF for networking~\cite{bertrone2018accelerating, tu2021revisiting, 9340283, 286467, 10.1145/3544216.3544259}.

ONCache includes the following components, as shown in Figure~\ref{fig: architecture}: (1) four eBPF programs for packet processing and forwarding, with their hook points listed in Table~\ref{tab:hook}; (2) three eBPF maps per host serving as the egress cache, the ingress cache, and the filter cache; (3) one user space daemon. The implementation of ONCache are much simpler compared to Slim and Falcon. The core functionalities can be realized with only 524 lines of eBPF C code, whereas Slim requires 2380 lines of code. Falcon, implemented with 606 lines of kernel code, is harder to implement and debug than eBPF. The core source code is shown in Appendix~\ref{app:code}.

Our idea diverges fundamentally from other cache-based network techniques. Prior works, such as OVS~\cite{15-ovs, rashelbach2022scaling} and Andromeda~\cite{dalton2018andromeda}, leverage caches to accelerate flow/rule matching on their data paths. However, as analyzed in \S\ref{sec:overheadana}, the container overlay network remains suboptimal even with the utilization of the OVS cache, revealing the insufficiency of caching results of a individual layer. In contrast, ONCache is the first \emph{cross-layer} cache within the container overlay network, which considers the extra overhead as a whole and caches the results once. Consequently, ONCache effectively eliminates the extra overhead compared to prior works.

%% file: sections/4Design.tex
\subsection{The Cache Definition}
\label{sec:cache}
ONCache contains three caches on each host, as shown in Figure~\ref{fig: architecture}: the egress cache, the ingress cache, and the filter cache.
These caches are implemented using eBPF Least Recently Used (LRU) hash map, which evicts the least recently used elements to store new elements when the map is full. 

The egress cache stores the outer headers and egress intra-host routing results, which are invariant for packets destined to the same container IP address. 
The routing results include the new inner MAC header (for L3 routing) and the egress host interface. To reduce the memory usage of the egress cache, we divide it into two levels: \emph{<container destination IP address (dIP) $\rightarrow$ host dIP>} and \emph{<host dIP $\rightarrow$ outer headers, inner MAC header, host interface index>}.

The ingress cache stores the ingress intra-host routing results, which are invariant for packets destined to the same container IP address. The structure of the ingress cache is \emph{<container dIP $\rightarrow$ inner MAC header, veth (host-side) index>}.

The filter cache stores the filtering decision of each flow in the established state.
Essentially, it functions as a flow whitelist, recording the allowed flows.
By default, a flow is defined by the 5-tuple (source IP address, source port, destination IP address, destination port, and transport protocol). One may also adjust the flow definition as required, e.g., adding a DSCP field to support DSCP filters.

In order to minimize cache misses, the capacity of the caches should be adjusted according to the cluster scale and concurrent flow number. Consider the largest cluster in Kubernetes~\cite{scale}, which has a maximum of 110 containers per host, 5k hosts, and 150k total containers. With up to 1M concurrent flows per host, to minimize cache misses, the egress/ingress/filter cache will take up at most 1.56 MB/2.2 KB/20 MB of memory space per host, respectively (See Appendix~\ref{app:mapsize} for details). This memory usage is negligible in modern servers.

\subsection{Cache Initialization}
\label{sec:cachemiss}
If ONCache encounters a cache miss while trying to forward a packet, it will pass the packet to the fallback overlay network and try to initialize the cache meanwhile. 
The cache initialization process is depicted in Figure~\ref{fig:init}. 

\textbf{Initialize the Egress Path.}
When an egress container packet reaches the veth (host-side), Egress-Prog, which is hooked here, tries to forward the packet. 
If either the egress cache or the filter cache misses for the packet, Egress-Prog adds a \emph{miss} mark to the packet.
We reserve one bit within DSCP field of the inner IP header as the \emph{miss} mark.\footnote{We reserve 2 bits in total within the inner IP headers' DSCP field and require the network not to use these 2 bits for differentiated services.}
Egress-Prog then passes the packet to the fallback overlay network.

The fallback overlay network proceeds to forward, filter, and encapsulate the packet. Besides, we configure the overlay network to add an \emph{est} mark to the packet if its flow reaches the established state. We use another bit in the DSCP field as the \emph{est} mark. 
The aforementioned action only requires either minor adjustments to two OVS flow rules or the addition of one rule in netfilter, as detailed in Appendix~\ref{app:cacheinit}. 

After being processed by the fallback overlay network, the packet reaches the host interface, where Egress-Init-Prog checks if the following requirements are met to initialize the caches: (1) the packet is a tunneling packet (e.g., a VXLAN packet); (2) the packet has the \emph{miss} mark and the \emph{est} mark. 

If all requirements are met, Egress-Init-Prog initializes the egress cache by storing \emph{<container dIP $\rightarrow$ outer headers, inner MAC header, host interface index>} into the cache. ONCache derives the host interface index from \verb|sk_buff| struct, and the other fields from the tunneling packet. Additionally, the flow is whitelisted in the filter cache. Finally, the packet is sent out from the host interface.

\textbf{Initialize the Ingress Path.}
Upon receiving a tunneling packet at the host interface, Ingress-Prog, which is hooked here, queries the cache and tries to forward the packet. If either the ingress cache or the filter cache misses for the packet, Ingress-Prog adds the \emph{miss} mark to the packet and passes it to the fallback overlay network. 

The fallback overlay network proceeds to process the packet, and add the \emph{est} mark to the packet if its flow reaches the established state.

Then the packet reaches the veth (container-side), where Ingress-Init-Prog checks if the packet has both the \emph{miss} and the \emph{est} mark. If so, the program initializes the cache by storing \emph{<container dIP $\rightarrow$ inner MAC header>} to the ingress cache and whitelisting the flow in the filter cache. \emph{<container dIP $\rightarrow$ veth (host-side) index>} in the ingress cache is maintained by ONCache daemon upon container provisioning.

\subsection{Cache-based Fast Path}
\label{sec:cachetrans}
We present the cache-based fast path of ONCache (Figure~\ref{fig:packet_journey}). 
The fast path is designed to be fail-safe and transparent to both applications and the underlay network. For packets that cannot be forwarded by the fast path, ONCache passes them to the fallback overlay network instead of dropping them.

\subsubsection{The Egress Fast Path}
\label{sec:egressdatapath}
The egress fast path is implemented by Egress-Prog, which handles an egress container packet in the following steps:

\textbf{Step \#1: Cache retrieving.} Egress-Prog first checks if the packet's flow is whitelisted in the filter cache and retrieves \emph{<container dIP $\rightarrow$ outer headers, inner MAC header, host interface index>} from the egress cache. Any cache miss triggers a cache initialization process, as described in \S \ref{sec:cachemiss}.

Additionally, Egress-Prog conducts a \emph{reverse check} by verifying whether the container source IP address exists in the ingress cache and if the reverse flow is also whitelisted in the filter cache. If the reverse check is not satisfied, Egress-Prog passes the packet to the fallback overlay network. 
This is necessary because for a flow, the eviction of caches for two directions and the expiration of its conntrack entry are asynchronous. 
Utilizing the fast path only when caches for both directions are ready ensures that if any cache initialization is needed, conntrack can observe traffic in both directions and add \emph{est} mark to packets, thereby ensuring the success of the initialization.
See a counterexample in Appendix~\ref{app:comp_conntrack}.

\textbf{Step \#2: Encapsulating and intra-host routing.}
Having retrieved the egress cache, Egress-Prog proceeds to rewrite the inner MAC header and add the outer headers to the packet. Two modifications on the outer headers are needed: (1) Updating the length, ID, and checksum fields in the outer IP header, as well as the length field in the outer UDP header; (2) Calculating the outer UDP source port using the same hash function employed by the kernel.
Finally, Egress-Prog invokes eBPF helper function \verb|bpf_redirect| to redirect the packet to the cached host interface index.

\subsubsection{The Ingress Fast Path}
The ingress fast path is implemented by Ingress-Prog, which handles an ingress tunneling packet in the following steps:

\textbf{Step \#1: Destination check.} 
Ingress-Prog first verifies whether destination MAC and IP addresses of the packet match with those of current host interface. TTL is also checked. If the destination check fails, the packet is then passed to the fallback overlay network for final decision.

\textbf{Step \#2: Cache retrieving.} Then, Ingress-Prog checks if the packet's flow is whitelisted in the filter cache, and retrieves \emph{<container dIP $\rightarrow$ inner MAC header, veth (host-side) index>} from the ingress cache. Any cache miss triggers a cache initialization process. Besides, the reverse check (described in \S \ref{sec:egressdatapath}) is also conducted on the ingress path.

\textbf{Step \#3: Decapsulating and intra-host routing.} 
Ingress-Prog proceeds to strip the outer headers of the packet, rewrite the inner MAC header, and redirect the packet to the destination veth (container-side). The redirection is done by invoking the eBPF helper function \verb|bpf_redirect_peer|.

Although the decapsulation of the outer headers is simplified in ONCache, no functionality is lost. Firstly, the payload is protected by checksums of the inner headers, which are verified by the container network stack. Secondly, the reassembly of fragmented packets is conducted by Generic Receive Offload (GRO) before reaching Ingress-Prog. Moreover, no other functionalities rely on the outer headers.

\subsection{Cache Coherency}
\label{sec:cachecoherency}
An overlay network is subject to change, necessitating ONCache to ensure cache coherency when changes occur.
While container provisioning is inherently handled by cache initialization, ONCache handles other changes by the user space program, ONCache daemon. 

Upon container deletion or unexpected container failures, ONCache daemon deletes the related caches. This prevents a new container with an old IP address from mistakenly utilizing outdated cache entries.

Upon other changes, including container migration or filter updates, ONCache employs a \emph{delete-and-reinitialize} mechanism with four steps to ensure that the changes take effect immediately and properly in the fast path:
(1) Pausing cache initialization by disabling OVS or netfilter from adding the \emph{est} mark. 
(2) Removing the cache entries affected by the network change (the affected packets start using the fallback overlay network).
(3) Applying the network change in the fallback overlay network 
(the network change takes effect immediately).
(4) Resuming cache initialization (the cache entries are reinitialized and the affected packets start using the fast path again).

\subsection{Cache Compatibility}
\label{sec:compatibilityconsiderations}
\textbf{Work with various traffic.} ONCache is designed to accelerate inter-host container traffic and is not responsible for other types of traffic, such as intra-host container traffic, container-to-host-IP traffic, or container-to-external-IP traffic. These types of traffic are handled by the fallback overlay network. 

The ClusterIP Service~\cite{service} in Kubernetes facilitates container access to a group of containers providing the same functionality via a single ClusterIP. 
Typically, such services rely on netfilter or IPVS (IP Virtual Server)~\cite{ipvs} on the sender host for load balancing and Destination Network Address Translation (DNAT). However, ONCache's fast path bypasses netfilter and IPVS. Nevertheless, ONCache can support ClusterIP akin to Cilium's approach: implementing load balancing and DNAT by eBPF programs and maps~\cite{ciliumclusterip}. This functionality can be integrated in Egress/Ingress-Prog and be compatible with the cache-based fast path.

\textbf{Work with data-plane policies.} 
Data-plane policies, such as rate limiting and quality of service (QoS) seamlessly integrate with ONCache. In Linux, these policies are implemented by queuing disciplines (qdiscs). ONCache's fast path does not bypass the qdiscs of the host interface.

\textbf{Filters not supported by ONCache.} Packet-based filters that match on the hash of the overlay packet~\cite{zhuo2019slim} do not exhibit any invariance property. 
Besides, while all stateful filters in netfilter~\cite{ayuso2006netfilter}, OVS~\cite{ovsconntrack}, Cilium~\cite{ciliumconntrack} have an established state, there may exist stateful filters without an established state. 
ONCache does not natively support such filters, because there is no invariant filter decision that can be cached. Nonetheless, it is feasible to implement such filters in eBPF case by case. 

\textbf{Work with service meshes.} Service meshes are increasingly popular in container clusters such as Kubernetes, primarily deployed for functionalities like security, observability, and traffic management~\cite{servicemesh}. The most common implementations of service meshes employ software proxies called sidecars~\cite{10.1145/3620678.3624652}. 
A sidecar is a separate process co-located with applications within the application network namespace (e.g., within a Kubernetes Pod~\cite{k8ssidecar}), and still relies on the overlay network for communication. Hence, ONCache benefits the communication of sidecar service meshes.

\textbf{Compatibility with Container Network Interface (CNI).} The current implementation of ONCache is based on TC eBPF. For CNIs that do not use TC eBPF in their data path, such as Antrea, Flannel, etc., ONCache can be seamlessly integrated without requiring any code modifications. Conversely, to use ONCache with CNIs that already utilize TC eBPF, such as Cilium, Calico, we need to reimplement ONCache within either the CNI's eBPF programs or the kernel source code.

\textbf{Container live migration.} 
ONCache is compatible with container live migration~\cite{criu, mirkin2008containers}. With proper control plane operations (see \S \ref{sec:cachecoherency}), the container connections can be well-maintained. 
In contrast, Slim only supports cold migration~\cite{zhuo2019slim}, as its container connections are established on the host and become invalid after migration, disrupting the applications inside the containers.

\textbf{Network debugging.} ONCache supports ICMP, thereby enabling common debugging tools such as ping and traceroute. Users can also utilize tools like bpftool~\cite{bpftool} to debug ONCache's eBPF programs and maps. Debugging with ONCache is easy and convenient. In contrast, debugging with Slim is much harder. Slim lacks support for ICMP, and there also lacks debug tools for its socket replacement mechanism.

Moreover, ONCache is compatible with Linux techniques such as segmentation offload~\cite{segmentation-offloads}, checksum offload~\cite{checksum-offloads}, and scaling techniques~\cite{rfs}.
See Appendix~\ref{app:compatibilitydiscussions} for details.

\subsection{Optional Improvements}
\label{sec:optimp}
We introduce two optional improvements for the fast path. Unlike the default design of ONCache, these optional improvements require either kernel or protocol modifications.

\begin{figure}[t]
    \centering
    \includegraphics[width=0.32\textwidth]{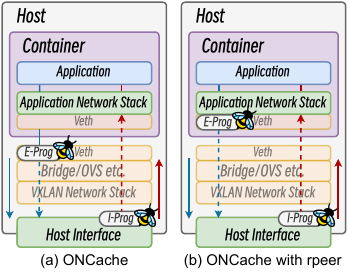}
    \setlength{\abovecaptionskip}{-5pt}
    \setlength{\belowcaptionskip}{-12pt}
    \caption{\textit{The fast path in (a) ONCache without rpeer; (b) ONCache with rpeer. 
    }}
    \label{fig:packet_journey_cmp}
\end{figure}

\textbf{Optimizing the redirect data path.}
ONCache uses \verb|bpf_redirect| and \verb|bpf_redirect_peer| to redirect packets on the egress and ingress path, respectively. As shown in Figure~\ref{fig:packet_journey_cmp}~(a), the two redirect functions are asymmetric, and the egress redirect path fails to eliminate the namespaces traversing overhead. 
Our speculation is that a symmetric egress redirect (Figure~\ref{fig:packet_journey_cmp}~(b)) requires hooking networking eBPF programs in container namespaces, which may conflict with user applications. Nevertheless, we are still curious about the potential benefit of optimizing the egress redirect data path.

We design and implement a new eBPF redirect function in the Linux kernel, called \verb|bpf_redirect_rpeer|. It provides a reversed path compared to \verb|bpf_redirect_peer|, redirecting packets from the egress port of the veth (container-side) to the egress port of the host interface, as depicted in Figure~\ref{fig:packet_journey_cmp}~(b).
By leveraging \verb|bpf_redirect_rpeer|, the hook point of Egress-Prog changes to TC egress of the veth (container-side), enabling ONCache to further mitigate the namespaces traversing overhead on the egress path. 

\textbf{Optimizing the tunneling protocol.}
Traditional tunneling protocols like VXLAN introduce transmission overhead due to the outer headers, typically tens of bytes (e.g., 50 bytes for VXLAN). To eliminate this transmission overhead, we propose and implement a rewriting-based tunneling protocol. 

With the new tunneling protocol, the fast path of ONCache works as below:
(1) Egress-Prog modifies container source/destination IP/MAC addresses of an egress container packet to the host ones and writes a \emph{restore key} to an idle field (e.g., DSCP) of the packet. 
Then the packet is redirected to the underlay network.
(2) Ingress-Prog restores all the addresses to their original values according to the restore key and redirect the packet to the destination container. 
The mapping between the restore key and the container source/destination IP/MAC addresses is cached during cache initialization.
See Appendix \ref{app:tunnelingproto} for more detail.

%% file: sections/6Evaluation.tex
\section{Evaluation}
\label{sec:evaluation}
\begin{figure*}[t]
    \centering
    \includegraphics[width=0.9\textwidth]{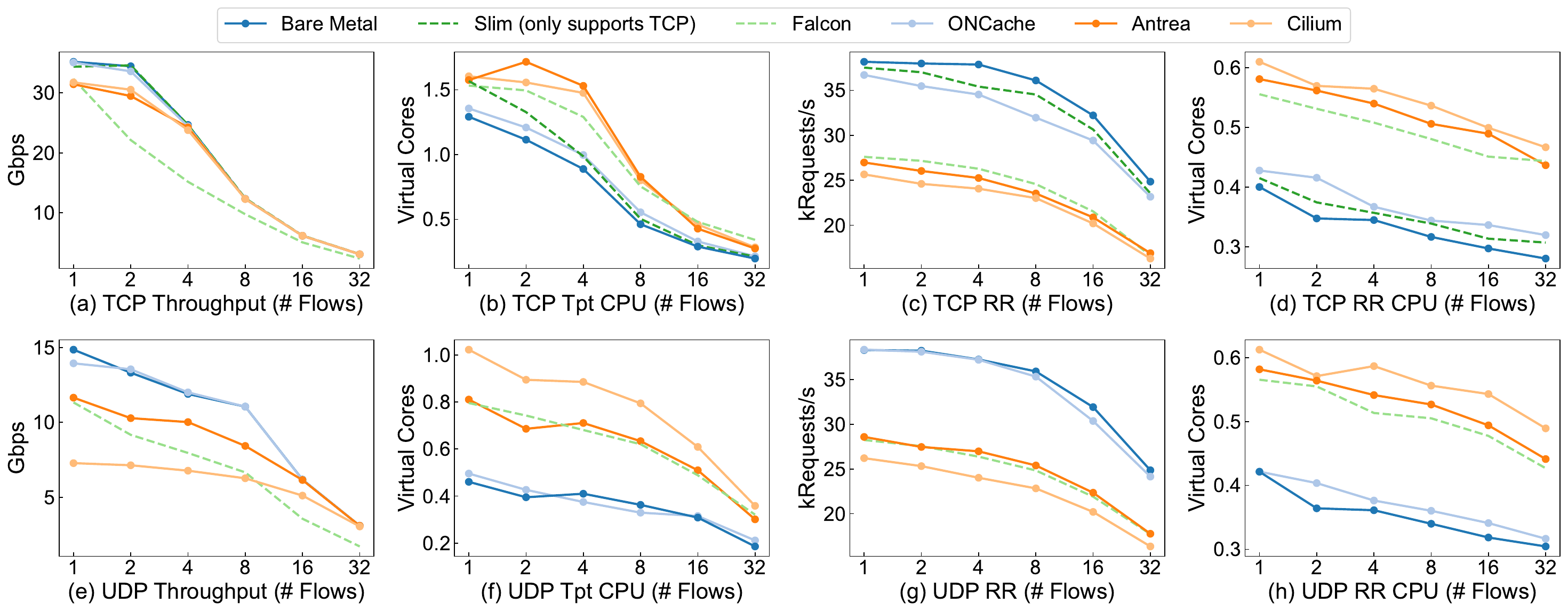}
    \setlength{\abovecaptionskip}{0pt}
    \setlength{\belowcaptionskip}{-10pt}
\caption{\textit{TCP and UDP microbenchmark results of bare metal, Slim (only supports TCP), Falcon (Linux kernel v5.4), ONCache, Antrea and Cilium. Both Cilium and Antrea provide standard overlay networks. All data is the average of a single flow. CPU utilization is measured on the receiver host, normalized by throughput or RR, and scaled to Antrea's throughput or RR.
}
}
    \label{fig:mb}
\end{figure*}

\begin{figure}
    \centering
    \includegraphics[width=0.45\textwidth]{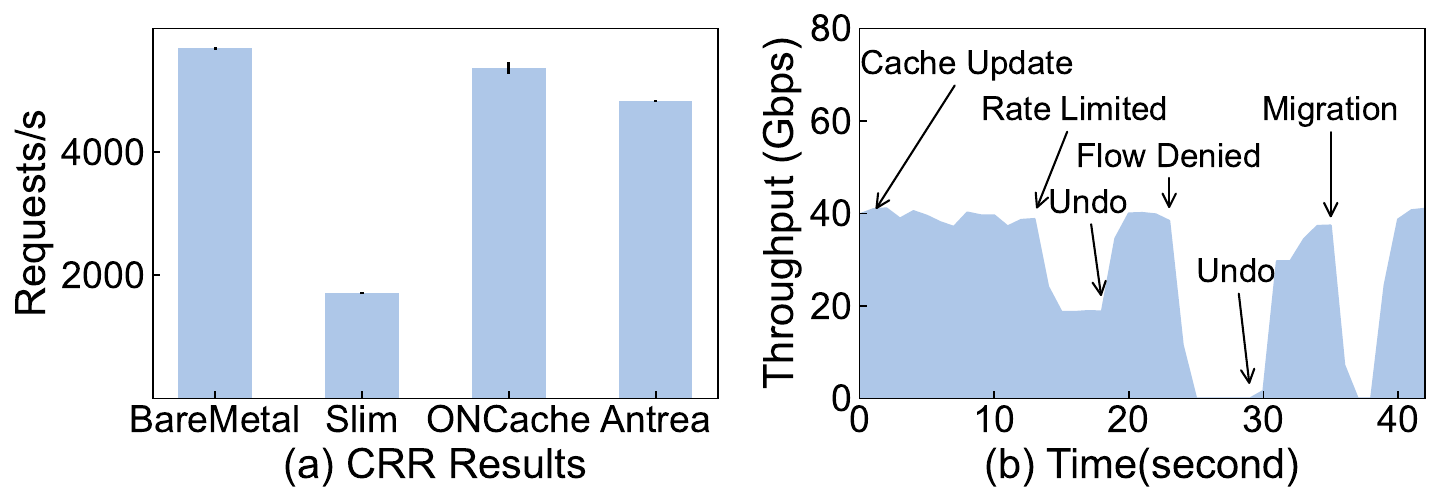}
    \setlength{\abovecaptionskip}{-2pt}
    \setlength{\belowcaptionskip}{-10pt}
    \caption{\textit{
    (a) The Connect-Request-Response (CRR) rate of different networks. The higher the better. Error bars denote standard deviations. (b) iperf3 throughput for the functional completeness experiments.}}
    
    \label{fig:conn}
\end{figure}

\begin{figure*}
    \centering
    \includegraphics[width=0.9\textwidth]{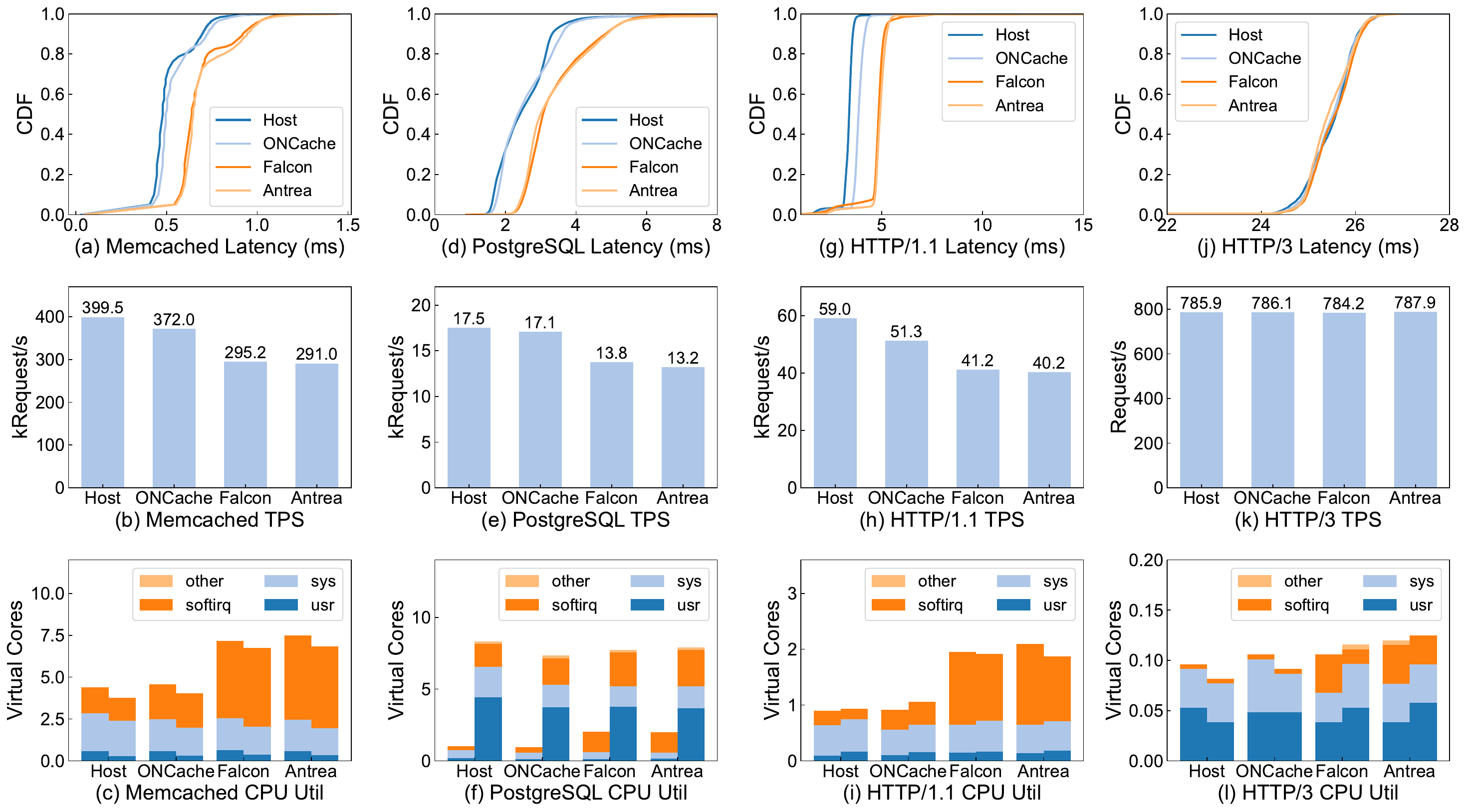}
    \setlength{\abovecaptionskip}{-5pt}
    \setlength{\belowcaptionskip}{-10pt}
\caption{\textit{The evaluation results of applications. Figures in each row present: the latency, the transaction rate (TPS) of all clients, and the CPU utilization of the client (left bar) \& the server (right bar). To ensure a fair comparison, we normalize the CPU utilization by TPS and scale it to Antrea's TPS. Softirq stands for software interrupt request.}
}
    \label{fig:app}
\end{figure*}

\begin{figure*}[t]
    \centering
    \includegraphics[width=0.9\textwidth]{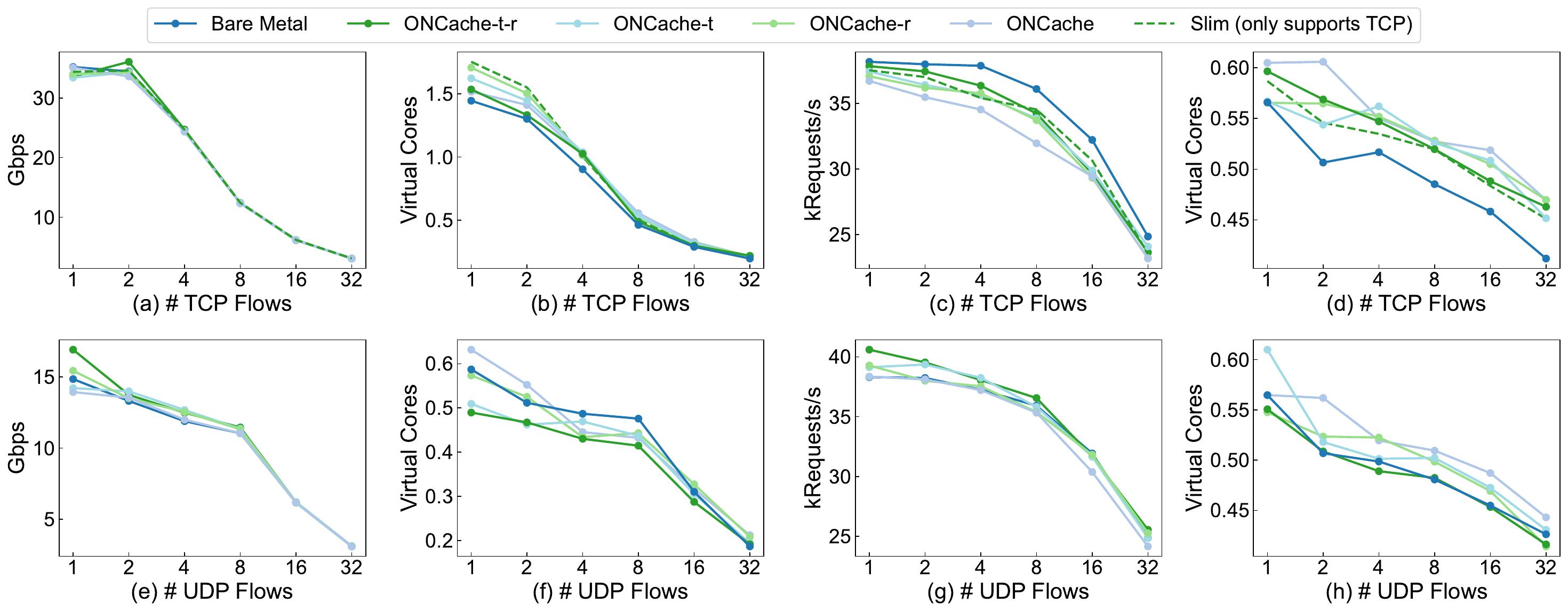}
    \setlength{\abovecaptionskip}{-5pt}
    \setlength{\belowcaptionskip}{-12pt}
\caption{\textit{TCP and UDP microbenchmark results of bare metal, ONCache with redirect rpeer (ONCache-r), ONCache with rewriting-based tunneling protocol (ONCache-t), ONCache with both (ONCache-t-r), and ONCache with neither (ONCache). 
The CPU utilization is normalized by throughput or RR and scaled to bare metal's throughput or RR.}
}
    \label{fig:mb_tr}
\end{figure*}

We evaluate ONCache (without optional improvements by default) with microbenchmarks in \S\ref{sec:mbm} and popular distributed applications in \S\ref{sec:app}. 
In both sections, ONCache's performance can be further improved by the optional improvements, at the cost of kernel/protocol modification. 
ONCache with optional improvements is evaluated in \S\ref{sec:optionalexp}.

Our testbed is built on top of a base Kubernetes~\cite{k8s} cluster (v1.23.6), including API server, placement engine, etcd, etc. We run the experiments on Cloudlab~\cite{Duplyakin+:ATC19} with three c6525-100g nodes, each equipped with an AMD EPYC 7402P 24-core processor@2.80 GHz. The processor uses hyperthreading. Each machine has 128 GB ECC memory and a dual-port Mellanox ConnectX-5 Ex 100 Gb interface. 
We do not change Cloudlab's default configurations of segmentation offload~\cite{segmentation-offloads}, checksum offload~\cite{checksum-offloads}, and scaling techniques~\cite{rfs}.
We use Ubuntu 20.04 with Linux kernel version 5.14. 
ONCache is deployed as a plugin of the Antrea (encap mode).

\subsection{Microbenchmarks}
\label{sec:mbm}
\subsubsection{Throughput and Latency}
We utilize iperf3~\cite{iperf3} to measure throughput and netperf~\cite{netperf} to measure Request-Response (RR) transaction rate. The RR test measures the rate of one-byte round-trips performed sequentially over a connection, wherein a higher transaction rate indicates lower end-to-end latency. We measure CPU utilization on the receiver host in all experiments using mpstat~\cite{mpstat}. Both TCP and UDP tests are executed.

In all settings, we conduct parallel tests. Multiple container pairs are deployed on a pair of hosts, with all server containers residing on one host and all client containers on the other. Each container pair starts testing simultaneously, enabling the evaluation of ONCache's parallel performance.

To represent performance of standard overlay networks, which serves as baseline, we select the most widely used CNIs, Antrea~\cite{antrea} and Cilium~\cite{cilium}.
We take bare metal as upper bound for overlay networks. 
To compare ONCache with prior works discussed in \S \ref{sec:relatedworks}, we conduct microbenchmarks on Slim\footnote{Slim does not support UDP, only the TCP microbenchmark is performed.} and Falcon, which represent the ideas of socket replacement and CPU load balancing, respectively.

\textbf{TCP microbenchmarks.} ONCache significantly improves both throughput and RR performance while consuming less CPU resource. The per-flow throughput, RR, and CPU utilization results are shown in Figure~\ref{fig:mb}~(a) – (d). 

Compared to Antrea, ONCache improves TCP throughput by 11.53\% and 13.96\% in 1 and 2-parallel tests. In 4, 8, 16, and 32-parallel tests, all container networks saturate the 100 Gb physical network. ONCache reduces per-byte CPU utilization by 13.94\% - 34.87\% in different parallel tests. 
The throughput and CPU utilization of ONCache are very close to those of Slim and bare metal.
The throughput of Falcon (only provides implementation in the kernel v5.4) is significantly lower. The reason is that the kernel v5.4 inherently exhibits lower bandwidth compared to the kernel v5.14 in our testbed.

RR performance of ONCache in all parallel tests outperforms Antrea by 35.81\% to 40.91\%, with a decrease in per-RR CPU utilization ranging from 26.02\% to 32.03\%. 
The RR results of ONCache exhibit a slight gap to those of Slim. Overhead, including egress namespace traversing, eBPF execution, and outer headers MTU overhead, contributes to this gap and can be mitigated by the optional improvements proposed in \S \ref{sec:optimp}.
As the RR test does not overwhelm any CPU cores, Falcon only slightly improves the RR results\footnote{In our testbed, the kernel version does not affect RR results.}.

\textbf{UDP microbenchmarks.} 
In contrast to Slim, ONCache also benefits UDP traffic, as shown in Figure~\ref{fig:mb} (e) – (h). 
UDP throughput of ONCache is 19.68\% - 31.76\% higher than Antrea in 1, 2, 4, 8-parallel tests. In 16 and 32-parallel tests, the throughput remains the same due to bandwidth bottleneck. Moreover, per-byte CPU utilization is lower by 29.73\% - 47.98\% in all tests. The throughput gap between ONCache and bare metal is less than 6\%. 

ONCache improves UDP RR performance by 34.13\% - 39.12\% in all tests, with per-RR CPU utilization reduced by 27.54\% - 31.59\%. There is also a small gap to Slim caused by the same reasons as in the TCP RR experiments. 

\textbf{Overhead quantifying.} We quantify ONCache's overhead by the method proposed in \S\ref{sec:overheadana} and present the results in Table~\ref{tab:overhead}.
As expected, ONCache eliminates all the extra overhead, except for egress namespaces traversing overhead, which is addressed by \verb|bpf_redirect_rpeer| proposed in \S\ref{sec:optimp}. Although incurring eBPF execution overhead, ONCache still contributes to a significant reduction in overall network stack latency. The quantitative analysis aligns consistently with the microbenchmark results previously presented.

\subsubsection{Cache Overhead}
\textbf{Cache initialization.}
We use netperf's Connect-Request-Response (CRR) test to measure TCP connection setup time and show the cache initialization overhead. In CRR, each RR transaction needs a new TCP connection, which requires additional cache initialization.

The CRR results are shown in Figure~\ref{fig:conn}~(a). ONCache is better than Antrea but worse than bare metal. This is because ONCache relies on Antrea to handle the first 3 packets before caches are initialized. ONCache performs the same as Antrea in this part. Then the round trip of 1 byte (Request-Response part) goes through the fast path, which performs better than Antrea. 
Falcon does not benefit CRR compared to Antrea due to the same reason as in TCP RR. 
Slim performs significantly worse. The reason is that Slim needs to first establish an overlay connection for service
discovery, which incurs several extra RTTs. 

\textbf{Cache interference.} We conduct an iperf3 test concurrently with continuous cache entry updates to show potential cache interference overhead. In this experiment, all cache capacities are set to 512, and the cache replacement policy is LRU. We use a script to continually insert 1000 redundant cache entries to the egress cache and subsequently delete them for 2 rounds. The test consumes about 8 seconds. As shown in the first 8 seconds of Figure~\ref{fig:conn}~(b), the iperf3 flow exhibits no significant throughput fluctuation.

\textbf{Cache scalability.} We conduct a TCP RR test with a full egress cache containing 150,000 entries (for the largest Kubernetes cluster as mentioned in \S\ref{sec:cache}). As expected, the RR performance remains unaffected, showcasing the inherent scalability of hash maps.

\subsubsection{Functional Completeness}
\label{sec:funccomplete}
\textbf{Data-plane policies.} 
We test whether data-plane policies, such as rate limiting, function correctly with ONCache. As shown in Figure~\ref{fig:conn}~(b), in the absence of rate limiting, the throughput reaches about 39 Gbps. Then we limit the rate to 20 Gbps on the host interface and the throughput drops to about 18.5 Gbps. Upon removal of the rate limit, the throughput returns to its original level.

\textbf{Packet filters.} We test whether packet filters work properly with ONCache. 
During an iperf3 test with ONCache, we apply a simple packet filter which denies this iperf3 flow using the delete-and-reinitialize mechanism proposed in \S \ref{sec:cachecoherency}. 
As expected, the throughput drops to 0, as shown in Figure~\ref{fig:conn}~(b). The throughput recovers upon removing the filter. 

\textbf{Container live migration.} 
We evaluate whether ONCache works properly with container live migration. Since Kubernetes does not natively support live migration, we imitate it by modifying the host IP address and VXLAN tunnels while the container remains alive.
As shown in Figure~\ref{fig:conn}~(b), the throughput drops to 0 when migration starts (the host IP address is changed) and recovers once the migration finishes (VXLAN tunnels are updated) after about 2 seconds.

Both the packet filter and container live migration experiments induce network changes and require using the delete-and-reinitialize mechanism (proposed in \S \ref{sec:cachecoherency}) to apply the changes. Thus these experiments also demonstrate the effectiveness of the delete-and-reinitialize mechanism.

\subsection{Applications}
\label{sec:app}
We evaluate ONCache on three real-world applications: in-memory key-value store Memcached~\cite{memcached}, database PostgreSQL~\cite{postgresql}, and web server Nginx~\cite{nginx}. We take Antrea as performance baseline and Docker host network~\cite{docker} as upper bound. 
Falcon is also evaluated. As Slim's performance is akin to that of the host network, it is omitted from this section. Notably, Slim only supports TCP, limiting its compatibility.

\textbf{Memcached.} We deploy a pair of containers on two hosts. One container runs a Memcached~\cite{memcached} server (v1.6), while the other runs a standard Memcached benchmark tool memtier~\cite{memtier} (v2.0.0) as a client. Upon the start of the experiment, the client spawns 4 threads, with each thread starting 50 connections. The ratio of SET and GET requests is 1:10. Memtier runs as fast as possible. We record CPU utilization by mpstat on both hosts during the experiment.

ONCache achieves much lower latency and higher transaction rate than Antrea. Figure~\ref{fig:app}~(a) is the CDF graph of the latency of GET requests. The average latency is reduced by 22.71\%, while the 99.9 percentile latency is reduced by 27.69\%. The latency gap relative to the host network is within 6\%. The transaction rates are shown in Figure~\ref{fig:app}~(b). ONCache outperforms Antrea by 27.83\%, while has a 7\% gap compared to the host network.

ONCache also significantly reduces per-transaction CPU utilization, as shown in Figure~\ref{fig:app}~(c). 
The per-transaction CPU utilization of ONCache is 38.91\% and 40.98\% lower on the client and the server, respectively. The utilization remains 4.09\% and 6.62\% higher than that in the host network.

\textbf{PostgreSQL.} We deploy a PostgreSQL database~\cite{postgresql} (v15.3) and its benchmark tool, pgbench~\cite{pgbench}, on a pair of containers on different hosts. Pgbench implements the TPC-B benchmark,  creating a database with 5 million banking accounts and executing 50 clients concurrently. Pgbench runs as fast as possible. We measure CPU utilization by mpstat.

ONCache benefits PostgreSQL's latency and transaction rate, as shown in Figure~\ref{fig:app}~(d) and (e). Compare to Antrea, ONCache reduces the average latency by 22.34\% and the transaction rate by 29.40\%, while having a gap of 2.30\% and 2.54\% to the host network, respectively.

ONCache also reduces CPU utilization, as shown in Figure~\ref{fig:app}~(f). The per-transaction CPU utilization is reduced by 51.84\% and 7.26\% for the client and the server, respectively. The reduction of the server is smaller due to relative high user CPU utilization. ONCache reduces soft interrupt utilization on the server by 27.36\%.

\textbf{Nginx.} To evaluate performance of HTTP servers, we conduct load tests on HTTP/1.1 and HTTP/3 (QUIC)~\cite{quic}, representing TCP-based and UDP-based HTTP, respectively. SSL is disabled for HTTP/1.1 and enabled for HTTP/3. We employ Nginx~\cite{nginx} (v1.25.1) and h2load~\cite{h2load} (v1.55.1) to benchmark the HTTP server. H2load runs 100 clients for HTTP/1.1 and 10 clients for HTTP/3. Each client issues 2 concurrent streams and requests a 1 KB file from the server. H2load runs as fast as possible.

ONCache improves latency and transaction rate for HTTP/1.1, as shown in Figure~\ref{fig:app}~(g) and (h). ONCache reduces HTTP/1.1 request latency by 21.53\% and improves the transaction rate by 27.43\%. There remains a performance gap between ONCache and the host network (15.04\% for latency and 13.12\% for transaction rate). The gap is caused by the same reason as analyzed in TCP RR, as the transaction rate is high enough to expose overhead. 
For HTTP/3, the performance is notably poorer and remains consistent across different networks, as shown in Figure~\ref{fig:app} (j) and (k). This could potentially be attributed to the experimental QUIC support of Nginx~\cite{expquic}.

ONCache reduces per-transaction CPU utilization for both HTTP/1.1 and HTTP/3, as shown in Figure~\ref{fig:app}~(i) and (l). For HTTP/1.1, CPU utilization is reduced by 56.26\% for the client and 43.52\% for the server. For HTTP/3, the reduction is 11.80\% and 26.76\% for the client and the server, respectively.

\subsection{Evaluation for Optional Improvements}
\label{sec:optionalexp}
We evaluate ONCache with redirect rpeer (ONCache-r), rewriting-based tunneling protocol (ONCache-t), and both of them (ONCache-t-r) by the methods described in \S \ref{sec:mbm}. The results are shown in Figure \ref{fig:mb_tr}. The baseline in this section is ONCache without the optional improvements.

The optional improvements benefit RR transaction rate, as shown in Figure~\ref{fig:mb_tr} (c)(g). ONCache-t, ONCache-r, and ONCache-t-r improve 1-parallel TCP RR performance by 1.96\%, 0.97\%, and 3.08\%, respectively (2.04\%, 2.43\%, and 5.87\% for UDP). 
For multi-parallel tests, the optional improvements also benefit RR performance. 
ONCache-t-r provides the most improvement in RR performance, nearly equaling the sum of those of ONCache-t and ONCache-r, and achieves nearly the same RR performance as Slim.

The optional improvements generally reduce CPU utilization, as shown in Figure~\ref{fig:mb_tr}~(b)(d)(f)(h). 
The CPU utilization reduction varies across different experiments, due to noise from other processes in the system.

We also evaluate the optional improvements with the applications used in \S \ref{sec:app}. The results are presented in Appendix~\ref{app:eval}.

%% file: sections/7Discussion.tex
\section{Discussion}
\label{sec:discussion}

\textbf{Why using TC hook?} There are two main types of eBPF hooks for networking: XDP and TC~\cite{BPF-reference}. Compared to XDP, TC eBPF programs do not require driver support, can redirect packets with lower overhead, are compatible with Linux traffic control (tc) module, and can be hooked on both ingress and egress ports. Therefore, TC is a better choice for ONCache.

\textbf{Security of eBPF-based ONCache.} 
Only processes that run by privileged user (root) or have the capability \verb|CAP_BPF| can load eBPF programs and read/write eBPF maps, unless the privileged user enables unprivileged eBPF~\cite{ebpfsafety}\footnote{Although there are userspace eBPF implementations~\cite{ubpf} that do not necessitate the privilege at all, they execute eBPF programs in userspace and differ from the eBPF adopted by ONCache.}. 
Thus, the eBPF components of ONCache are protected by the eBPF permission control. Additionally, ONCache's fast path resides in kernel mode and is inaccessible from applications. 
In contrast, Slim has serious security issues because it exposes host namespace file descriptors to containers, thereby breaking resource isolation provided by Linux namespaces. Slim's kernel module could mitigate the security issues, but resource isolation is still undermined.

%% file: sections/8Relatedworks.tex
\section{Related Works}
\textbf{Cache in network virtualization.} 
OVS~\cite{15-ovs, rashelbach2022scaling} employs a cache to accelerate flow matching. However, as analyzed in \S\ref{sec:overheadana}, there is still significant extra overhead in overlay networks.
Andromeda~\cite{dalton2018andromeda}, a VM cluster network virtualization stack, maintains a cache of routing and filtering decisions, sharing a similar idea with OVS.
Nevertheless, none of the prior works quantify container overlay networks and employ a cross-layer cache to bridge the performance gap between container overlay networks and bare metal, as ONCache does.

\textbf{eBPF-based container networks.} ONCache differs significantly from existing eBPF-based container networks like Cilium~\cite{cilium} and Calico~\cite{calico}. For example, in Cilium, the eBPF-based data path aims to replace netfilter in the host network stack with eBPF programs and is effective where container packets are directly forwarded to underlay network~\cite{ebpfhostrouting}. However, as analyzed in \S\ref{sec:overheadana}, Cilium fails to improve overlay networks and the extra overhead still exists. 
In contrast, ONCache aims to eliminate the extra overhead with the proposed cache, as shown in Table~\ref{tab:overhead}.

\textbf{Service mesh sidecar optimization.} 
While a sidecar simplifies traffic management, policy enforcement, network security, etc., it incurs significant overhead. Zhu et al. \cite{10.1145/3620678.3624652} proposed MeshInsight to quantify the overhead of service mesh sidecars, and works such as SPRIGHT~\cite{10.1145/3544216.3544259}, mRPC~\cite{chen2023remote} optimize sidecars by different mechanisms. However, these works focus on optimizing the overall intra-host service mesh architecture, rather than the inter-host container communication.

%% file: sections/9Conclusion.tex
\section{Conclusion}
We design and implement ONCache, a cache-based container overlay network that effectively narrows the performance gap between bare metal and overlay networks without losing flexibility and compatibility.
In both microbenchmark and application experiments, 
ONCache significantly outperforms standard overlay networks in terms of throughput and latency while reducing CPU utilization.
ONCache can be seamlessly integrated with existing CNIs such as Antrea, Flannel, etc.

\clearpage

%% file: sections/10appendix.tex
\clearpage
\appendix
\section{Quantitatively Analyzing Network Stacks}
\label{app:bcctiming}
In this section, we describe how we quantitatively analyze network stacks. We analyze the overhead incurred during the execution of a 1-byte TCP RR test conducted using NPtcp~\cite{netpipe}. 

Firstly, we perform a qualitative analysis of overhead in different data paths with two steps: (1) Utilizing the perf tool~\cite{perf}, we profile network stack functions for a duration of 10 seconds and subsequently employ a flame graph generator~\cite{flamegraph} to generate flame graphs. (2) According to the flame graphs, we classify network stack functions into different segments of the overhead, e.g., the application network stack, OVS, etc.

Next, we time the network stack functions using eBPF one by one. As the network stack functions may execute in both the application network stack and the VXLAN network stack, we need to distinguish them by their call stacks. We implement two eBPF programs leveraging the BPF Compiler Collection (BCC)~\cite{bcc} tool. These programs are hooked to \verb|kprobes| and \verb|kretprobes| hook points, executing at the entry and return points of each specific function, respectively. They capture the current call stack along with the timestamp, and store the function (with the call stack) and its execution time in an eBPF map. For each function, we record all the execution time samples within one second and compute the average of these samples, which serves as the final result, as shown in Table~\ref{tab:overhead}.

\section{The Core Code of ONCache}
\label{app:code}
In this section, we present core code of ONCache's eBPF programs to help to understand the design of ONCache. 
\subsection{The Local Cache Definition}
The code in this section implements functionalities that are described in \S \ref{sec:cache}.

\textbf{The egress cache:}
\begin{lstlisting}[language=C]
struct egressinfo {
    unsigned char outer_header[64];
    __u32 ifidx;
};
struct bpf_elf_map SEC("maps") egressip_cache = {
    .type = BPF_MAP_TYPE_LRU_HASH,
    .size_key = sizeof(__be32),
    .size_value = sizeof(__be32),
    .pinning    = PIN_GLOBAL_NS,
    .max_elem = 4096
};
struct bpf_elf_map SEC("maps") egress_cache = {
    .type = BPF_MAP_TYPE_LRU_HASH,
    .size_key = sizeof(__be32),
    .size_value = sizeof(struct egressinfo),
    .pinning    = PIN_GLOBAL_NS,
    .max_elem = 1024,
};
\end{lstlisting}

\textbf{The ingress cache.}
\begin{lstlisting}[language=C]
struct ingressinfo {
    __u32 ifidx;
    unsigned char dmac[ETH_ALEN];
    unsigned char smac[ETH_ALEN];
};
struct bpf_elf_map SEC("maps") ingress_cache = {
    .type = BPF_MAP_TYPE_LRU_HASH,
    .size_key = sizeof(__be32),
    .size_value = sizeof(struct ingressinfo),
    .pinning    = PIN_GLOBAL_NS,
    .max_elem = 1024,
};
\end{lstlisting}

\textbf{The filter cache.}
\begin{lstlisting}[language=C]
struct action {
    __u16 ingress;
    __u16 egress;
};
struct bpf_elf_map SEC("maps") filter_cache = {
    .type = BPF_MAP_TYPE_LRU_HASH,
    .size_key = sizeof(struct fivetuple),
    .size_value = sizeof(struct action),
    .pinning    = PIN_GLOBAL_NS,
    .max_elem = 4096,
};
\end{lstlisting}

\subsection{Cache Initialization}
\label{app:cacheinit}
The code in this section implements functionalities that are described in \S \ref{sec:cachemiss}.

\textbf{Initialize the Egress Path.} \verb|TC_ACT_OK| indicates the kernel to proceed with the usual packet processing. ONCache uses \verb|TC_ACT_OK| to pass the packet to the fallback overlay network.
\begin{lstlisting}[language=C]
// Checks if miss and est marked.
if ((inner_iph->tos & 0xc) != 0xc) return TC_ACT_OK;
// Update filter cache
struct fivetuple tuple_;
if (parse_5tuple_e(inner_iph, data_end, &tuple_)) return TC_ACT_OK;
struct action eaction_ = {
    .egress = 1,
    .ingress = 0
};
if(bpf_map_update_elem(&filter_cache, &tuple_, &eaction_, BPF_NOEXIST)) {
    struct action* action_ = bpf_map_lookup_elem(&filter_cache, &tuple_);
    if (action_) action_->egress = 1;
}
// Update egress cache
struct egressinfo egressinfo_;
initegressinfo(&egressinfo_, skb);
if(bpf_map_update_elem(&egress_cache, &outer_iph->daddr, &egressinfo_, BPF_NOEXIST))
    return TC_ACT_OK;
if(bpf_map_update_elem(&egressip_cache, &inner_iph->daddr, &outer_iph->daddr, BPF_NOEXIST))
    return TC_ACT_OK;
// Erase the TOS mark.
set_ip_tos(skb, 50, 0);
\end{lstlisting}

\textbf{Initialize the Ingress Path.}
\begin{lstlisting}[language=C]
// Checks if miss and est marked.
if ((iphdr->tos & 0xc) != 0xc) return TC_ACT_OK;
// Update ingress cache
struct ingressinfo* ingressinfo_ = bpf_map_lookup_elem(&ingress_cache, &iphdr->daddr);
if (!ingressinfo_) {
    return TC_ACT_OK;
} else {
    __builtin_memcpy(ingressinfo_->dmac, eth->h_dest, ETH_ALEN);
    __builtin_memcpy(ingressinfo_->smac, eth->h_source, ETH_ALEN);
}
// Update filter cache
struct fivetuple tuple_;
if (parse_5tuple_in(iphdr, data_end, &tuple_)) return TC_ACT_OK;
struct action eaction_ = {
    .egress = 0,
    .ingress = 1
};
if(bpf_map_update_elem(&filter_cache, &tuple_, &eaction_, BPF_NOEXIST)) {
    struct action* action_ = bpf_map_lookup_elem(&filter_cache, &tuple_);
    if (action_) action_->ingress = 1;
}
// Erase the TOS mark.
set_ip_tos(skb, 0, 0);
\end{lstlisting}

\begin{figure}[t]
    \centering
    \includegraphics[width=0.5\textwidth]{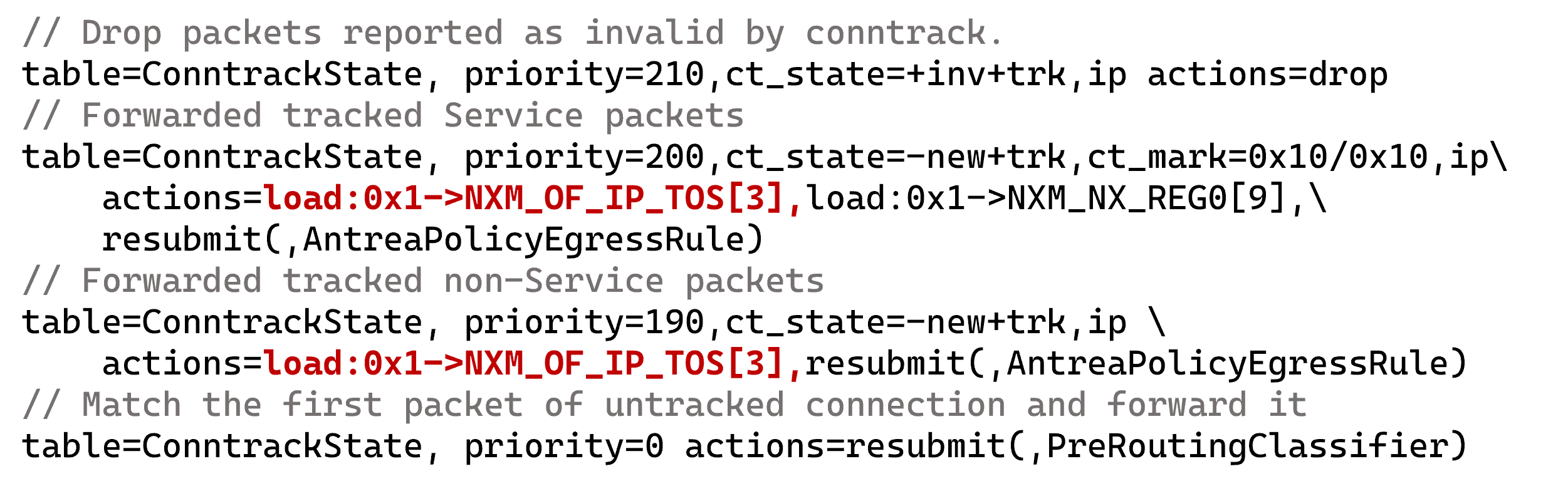}
    \caption{\textit{Parts of OVS flows in Antrea. The action that we introduce is highlighted in red font, through which we set a predefined DSCP bit to 1 if the flow reaches established state.}}
    \label{fig:ovsflow}
\end{figure}

\textbf{Extra configuration on OVS or netfilter.} 
The addition of the \emph{est} mark can be realized through either OVS or netfilter.

For OVS, this action requires modifying two OVS flows, as shown in Figure \ref{fig:ovsflow}. 
Originally, the two flows are intended for forwarding non-new-state tracked packets. We utilize them to add the \emph{est} mark to packets meanwhile. The actions that we introduce are highlighted in red font, through which we set a predefined DSCP bit to 1 if the flow reaches established state.

Alternatively, this action can be independently realized by adding a new rule in netfilter as follows.

\begin{lstlisting}[language=C]
iptables -t mangle -A FORWARD -m conntrack --ctstate ESTABLISHED -m dscp --dscp 0x1 -j DSCP --set-dscp 0x3
\end{lstlisting}

\subsection{Cache-based Fast Path}
The code in this section implements functionalities that are described in \S \ref{sec:cachetrans}.

\subsubsection{The Egress Data Path}
\textbf{Step \#1: Cache retrieving.} 
\begin{lstlisting}[language=C]
struct fivetuple tuple_;
if (parse_5tuple_e(iphdr, data_end, &tuple_)) return TC_ACT_OK;
struct action *action_ = bpf_map_lookup_elem(&filter_cache, &tuple_);
if (!action_ || !(action_->ingress & action_->egress)) {
    set_ip_tos(skb, 0, 0x4);
    return TC_ACT_OK;
}
__be32* nodeip_ = bpf_map_lookup_elem(&egressip_cache, &iphdr->daddr);
if (!nodeip_) {
    // TOS 0x4 is used as miss mark
    set_ip_tos(skb, 0, 0x4);
    return TC_ACT_OK;
} 
struct egressinfo* egressinfo_ = bpf_map_lookup_elem(&egress_cache, nodeip_);
if (!egressinfo_) {
    set_ip_tos(skb, 0, 0x4);
    return TC_ACT_OK;
}
struct ingressinfo* ingressinfo_ = bpf_map_lookup_elem(&ingress_cache, &iphdr->saddr);
if (!ingressinfo_ || !ingressinfo_complete(ingressinfo_)) 
    return TC_ACT_OK;
\end{lstlisting}

\textbf{Step \#2: Encapsulating and intra-host routing.} 
\begin{lstlisting}[language=C]
// Prepend the outer headers
if (bpf_skb_adjust_room(skb, 50, BPF_ADJ_ROOM_MAC, BPF_F_ADJ_ROOM_FIXED_GSO | BPF_F_ADJ_ROOM_ENCAP_L3_IPV4 | BPF_F_ADJ_ROOM_ENCAP_L4_UDP| BPF_F_ADJ_ROOM_ENCAP_L2(14)| BPF_F_ADJ_ROOM_ENCAP_L2_ETH)))
    return TC_ACT_OK;
// Boundary check
... 
// Headers includes 50B of outer hdr and 14B inner MAC hdr.
__builtin_memcpy(data, egressinfo_->headers, 64);
set_lengthandid(skb, skb->len);
// Set the UDP source port
__u32 hash = bpf_get_hash_recalc(skb);
__be16 sport = get_udpsport(hash);
bpf_skb_store_bytes(skb, UDP_PORT_OFF, &sport, sizeof(sport), 0);
// Packet redirecting
action = bpf_redirect(egressinfo_->ifidx, 0);
return action;
\end{lstlisting}

\subsubsection{The Ingress Data Path}
\textbf{Step \#1: Destination check.} 
\begin{lstlisting}[language=C]
struct bpf_elf_map SEC("maps") devmap = {
    .type = BPF_MAP_TYPE_HASH,
    .size_key = sizeof(int),
    .size_value = sizeof(struct devinfo),
    .pinning    = PIN_GLOBAL_NS,
    .max_elem = 8,
};

struct ethhdr *outer_eth = data;
int ifindex = ctx->ifindex;
struct devinfo *devinfo_ = bpf_map_lookup_elem(&devmap, &ifindex);
if (!devinfo_ || maccmp(outer_eth->h_dest, devinfo_->mac, ETH_ALEN)) return TC_ACT_OK;
// Check if Ethernet frame has IP packet and set IP hdr ptr
if (outer_eth->h_proto != bpf_htons(ETH_P_IP)) return TC_ACT_OK;
struct iphdr *outer_iph = (struct iphdr *)(outer_eth + 1);
if (outer_iph->daddr != devinfo_->ip) return TC_ACT_OK;
\end{lstlisting}

\textbf{Step \#2: Cache retrieving.} 
\begin{lstlisting}[language=C]
struct fivetuple tuple_;
if (parse_5tuple_in(inner_iph, data_end, &tuple_)) return TC_ACT_OK;
struct action *action_ = bpf_map_lookup_elem(&filter_cache, &tuple_);
if (!action_ || !(action_->ingress & action_->egress)) {
    set_ip_tos(skb, 50, 0x4);
    return TC_ACT_OK;
}
struct ingressinfo* ingressinfo_ = bpf_map_lookup_elem(&ingress_cache, &inner_iph->daddr);
if (!ingressinfo_ || !ingressinfo_complete(ingressinfo_)) {
    set_ip_tos(skb, 50, 0x4);
    return TC_ACT_OK;
}
if (!bpf_map_lookup_elem(&egressip_cache, &inner_iph->saddr)) return TC_ACT_OK;
\end{lstlisting}

\textbf{Step \#3: Decapsulating and intra-host routing.} 
\begin{lstlisting}[language=C]
if (bpf_skb_adjust_room(ctx, -50, BPF_ADJ_ROOM_MAC, 0)) return TC_ACT_OK;
// Boundary check
...
// Set new MAC
struct ethhdr *eth = data;
__builtin_memcpy(eth->h_dest, ingressinfo_->dmac, ETH_ALEN);
__builtin_memcpy(eth->h_source, ingressinfo_->smac, ETH_ALEN);
// Packet redirecting
action = bpf_redirect_peer(ingressinfo_->ifidx, 0);
return action;
\end{lstlisting}

\section{Calculation of the Maps Size}
\label{app:mapsize}
As per the cache definitions described in \S\ref{sec:cache}, the sizes of cache entries are specified as below: 8 bytes for the egress cache (first level), 72 bytes for the egress cache (second level), 20 bytes for the ingress cache, and 20 bytes for the filter cache. 

To eliminate cache eviction by the LRU mechanism for a cluster with a maximum of 110 containers per host, 5k hosts, 150k total containers, and up to 1M concurrent flows per host, we need the egress cache (first level) to have 150k entries, the egress cache (second level) to have 5k entries, the ingress cache to have 110 entries, and the filter cache to have 1M entries. 

Consequently, the sizes of the caches can be calculated as follows: the egress cache is 8 B * 150k + 72 B * 5k = 1.56 MB, the ingress cache is 20 * 110 = 2.2 KB, and the filter cache is 20 B * 1M = 20 MB.

\section{An Example to Understand Necessity of Reverse Check}
\label{app:comp_conntrack}
We further use a counterexample to show the necessity of reverse check. 

Let's consider a scenario where ONCache only checks the egress cache on the egress data path. Initially, flow $f$ enters the established state in conntrack, and all caches for $f$ are initialized. Consequently, all packets of flow $f$ go through ONCache's fast path. However, since flow $f$ bypasses the conntrack module in the fallback overlay network, its conntrack entry eventually expires after a predefined duration. Subsequently, if flow $f$ is evicted from the ingress cache by LRU mechanism, it can still leverage the egress fast path but becomes ineligible for the ingress fast path. Ideally, we envision ONCache to reinitialize the ingress cache for flow $f$. But unfortunately, flow $f$ cannot re-enter the established state in conntrack because conntrack records a flow as established only upon observing packets in both directions~\cite{ayuso2006netfilter}. Consequently, the ingress cache for flow $f$ can never be reinitialized and flow $f$ can no longer utilize the ingress fast path.

\section{Compatibility Discussions}
\label{app:compatibilitydiscussions}
\textbf{Segmentation offloads.} Segmentation offloads, such as Generic Segmentation Offload (GSO) and Generic Receive Offload (GRO)~\cite{segmentation-offloads}, enable network interfaces to segment and reassemble packets. The hooking points of all ONCache eBPF programs are TC. On the egress path, GSO happens after TC~\cite{BPF-reference}. 
Thus GSO and other hardware segmentation mechanisms of the host interface are not bypassed by the fast path. 
On the ingress path, GRO happens prior to TC~\cite{BPF-reference}, posing no conflicts with ONCache. Hence, ONCache is compatible with segmentation offload techniques.

\textbf{Scaling techniques.} Linux scaling techniques, including Receive Packet Steering (RPS) and Receive Flow Steering (RFS) (and their hardware counterparts Receive Side Scaling (RSS) and Accelerated RFS (aRFS))~\cite{rfs}, employ different CPU load balancing strategies on the ingress data path. They are all compatible with ONCache: RSS and aRFS are hardware mechanisms; RPS and RFS are in software but happen before Ingress-Prog on the ingress path.

\textbf{Checksum offloads.} 
Checksum offloads, including transmitting and receiving checksum offload~\cite{checksum-offloads}, are often enabled to reduce checksum overhead. As they are hardware mechanisms, ONCache is compatible with them.

\begin{figure}[t]
    \centering
    \includegraphics[width=0.45\textwidth]{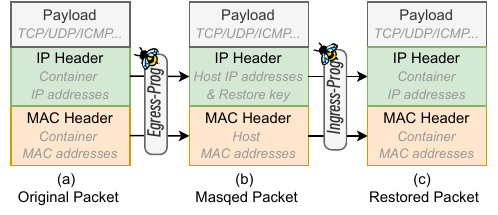}
    \caption{\textit{The changes of a container packet forwarded by ONCache with the rewriting-based tunneling protocol. (a) The original packet from containers. (b) The masqueraded packet. (c) The restored packet.
    }}
    \label{fig:packet_changing}
\end{figure}
\begin{figure*}[!t]
    \centering
    \includegraphics[width=0.82\textwidth]{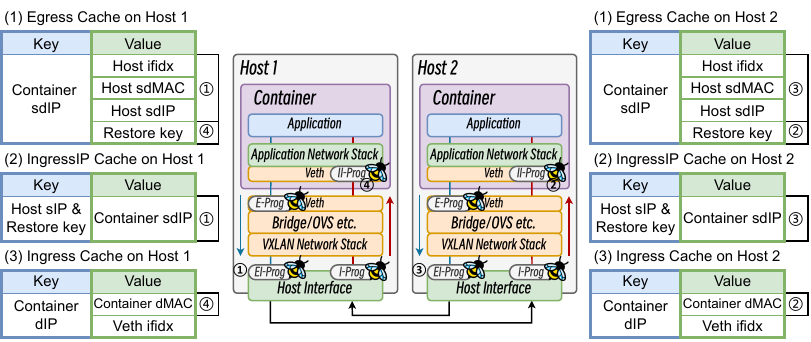}
\caption{\textit{The cache initialization of ONCache with the rewriting-based tunneling protocol. \ding{172} - \ding{175} mark four steps where parts of the caches are filled up. ifidx is short for interface index.}
}
    \label{fig:hdrinit}
\end{figure*}

\begin{table*}[t]
\centering
\begin{tabular}{c|l|ccccc}
\toprule
\multicolumn{2}{c|}{\textbf{Results}} & \textbf{ONCache-t} & \textbf{ONCache-r} & \textbf{ONCache-t-r} & \textbf{Host} & \textbf{ONCache} \\
\midrule
\multirow{3}{*}{Memcached}  & Latency & -2.94\% & -7.01\% & -6.04\% & -5.38\% & 0\% \\
                            & TPS     & +2.97\% & +1.74\% & +4.32\% & +7.42\% & 0\% \\
                            & CPU     & +1.92\% & +2.14\% & -1.10\% & -6.21\% & 0\% \\
\midrule
\multirow{3}{*}{PostgreSQL} & Latency & +0.04\% & -1.28\% & -5.77\% & -2.25\% & 0\% \\
                            & TPS     & -0.27\% & +1.23\% & +6.21\% & +2.61\% & 0\% \\
                            & CPU     & -1.86\% & -4.61\% & -10.36\% & +13.17\% & 0\% \\
\midrule
\multirow{3}{*}{HTTP/1.1}   & Latency & -2.82\% & -8.46\% & -10.00\% & -13.08\% & 0\% \\
                            & TPS     & +2.78\% & +9.09\% & +10.90\% & +15.10\% & 0\% \\
                            & CPU     & -5.48\% & -11.93\% & -14.02\% & -11.26\% & 0\% \\
\midrule
\multirow{3}{*}{HTTP/3}     & Latency & 0.00\% & +0.16\% & +0.16\% & +0.04\% & 0\% \\
                            & TPS     & -0.03\% & -0.03\% & -0.01\% & -0.03\% & 0\% \\
                            & CPU     & -26.30\% & +15.82\% & -10.51\% & -10.50\% & 0\% \\
\bottomrule
\end{tabular}
\caption{\textit{The application performance and server CPU utilization (normalized by TPS) of ONCache-t, ONCache-r, ONCache-t-r, and the host network compared to ONCache.}}
\label{tab:ONCachetr}   
\end{table*}
\section{Rewriting-based Tunneling Protocol}
\label{app:tunnelingproto}
This section describes how the rewriting-based tunneling protocol works in ONCache. 

The rewriting-based tunneling protocol operates by substituting the encapsulation of outer headers in traditional tunneling protocols, with the rewriting of inner headers. The modifications to packet headers along the data path are depicted in Figure~\ref{fig:packet_changing}.

\textbf{New caches.} The egress and ingress caches are redesigned to accommodate the rewriting-based tunneling protocol.

The egress cache stores data used on the egress path, structured as \emph{<container sdIP $\rightarrow$ host interface index, host sdIP, host sdMAC, restore key>}\footnote{sIP = source IP address; dIP = destination IP address. Similarly, for sMAC, dMAC, sdIP, and sdMAC.}.

The ingress cache stores data used on the ingress path, structured as \emph{<host sIP \& restore key $\rightarrow$ veth (host-side) index, container sdIP, container dMAC>}.
To accommodate the new initialization process described later, the ingress cache is divided into two maps. The first is the ingressIP cache: \emph{<host sIP \& restore key $\rightarrow$ container sdIP>}, while the second is the ingress cache: \emph{<container dIP $\rightarrow$ container dMAC, veth (host-side) index>}.

\textbf{The new packet journey.} On the sender host, Egress-Prog retrieves 
\emph{<container sdIP $\rightarrow$ host interface index, host sdIP, host sdMAC, restore key>} from the egress cache. Subsequently, Egress-Prog conducts the \emph{masquerading} process by modifying the container source/destination MAC/IP addresses to those of the host. 

To ensure the masqueraded packet can be restored on the receiver host, ONCache additionally writes a restore key to the packet. Then the receiver host utilizes the restore key in conjunction with the host IP addresses to restore container source/destination MAC/IP addresses. The restore key is allocated during cache initialization and is stored in the egress cache. The user has the flexibility to designate a specific field to accommodate the restore key within the packet. This field can be any available field in the IP header, such as ID, DSCP, or Option. The capacity of the restore key depends on the width of the chosen field. After writing the restore key to the packet, referred to as the masqueraded packet, Egress-Prog redirects it to the host interface using either \verb|bpf_redirect| or \verb|bpf_redirect_rpeer|.

On the receiver host, Ingress-Prog restores the masqueraded packet. Ingress-Prog retrieves \emph{<host sIP \& restore key $\rightarrow$ veth (host-side) index, container sdIP, container dMAC>}
from the ingress cache, and writes the container IP and MAC addresses back to the headers. After restoring, the packet is redirected to the destination veth using \verb|bpf_redirect_peer|.

\textbf{The new cache initialization.} The initialization for MAC and IP addresses, as well as interface indexes, remains akin to the original design (as described in \S \ref{sec:cachemiss}), and the allocation of the restore key is newly added. The restore key is allocated on the receiver host and subsequently delivered to the sender host. This ensures that the masquerading operation utilizes the restore key only after the receiver host recognizes it. The initialization of the filter cache is the same as that described in \S\ref{sec:cachemiss} and is therefore omitted in this section. ONCache utilizes a round-trip of tunneling packets to initialize the cache. The process is shown in Figure~\ref{fig:hdrinit} and encompasses four steps:

\emph{Step} \#1 (tagged in Figure~\ref{fig:hdrinit} \ding{172}):
Upon encountering a cache miss, Egress-Prog passes the packet to the fallback overlay network. Then the packet reaches Egress-Init-Prog after undergoing processing within the fallback overlay network. Egress-Init-Prog parses the headers and stores \emph{<container sdIP $\rightarrow$ host interface index, host sdIP, host sdMAC>} to the egress cache. It also allocates a restore key for the reverse flow, chosen randomly or sequentially, and writes \emph{<host sIP \& restore key $\rightarrow$ container sdIP>} to the ingressIP cache. As a hash map, the ingressIP cache naturely ensures the uniqueness of the restore key. To deliver the restore key to the peer host, Egress-Init-Prog writes the restore key into the user-defined field within the inner header. Finally, the packet is transmitted to the underlay network.

\emph{Step} \#2 (tagged in Figure \ref{fig:hdrinit} \ding{173}): Upon receiving a tunneling packet on the receiver host, Ingress-Init-Prog continues the initialization. It parses the restore key contained within the packet, and stores \emph{<container sdIP $\rightarrow$ restore key>} to the egress cache. Next, it stores \emph{<container dIP $\rightarrow$ container dMAC>} to the ingress cache (\emph{<container dIP $\rightarrow$ veth (host-side) index>} is maintained by the user space daemon). Finally, the packet is forwarded to the destination application.

The above two steps fill up half of fields within the egress cache and ingress cache on both hosts. Subsequently, the reply of this tunneling packet goes through a similar yet reverse process (tagged in Figure \ref{fig:hdrinit} \ding{174}\ding{175}), thereby completing the entire initialization process.

\section{Application Evaluation for Optional Improvements}
\label{app:eval}
This section, we show the application evaluation results for the optional improvements. The evaluation results of ONCache-r, ONCache-t, ONCache-t-r, and the host network are listed in Table~\ref{tab:ONCachetr}. The listed values are relative to ONCache.

The application evaluation reveals the following key findings: 
(1) Both the rewriting-based tunneling protocol (ONCache-t) and \verb|bpf_redirect_rpeer| (ONCache-r) improve latency and TPS performance across all applications, except for HTTP/3. They also reduce per-transaction CPU utilization for PostgreSQL and HTTP/1.1.
(2) With both optional improvements, ONCache-t-r generally achieves the best latency, TPS, and CPU utilization across all applications, except for HTTP/3. Notably, the performance of ONCache-t-r closely rivals that of the host network.

Due to the constraints of the experimental QUIC support of Nginx, the absolute result values of HTTP/3 across different implementations exhibit slight variations. Consequently, the evaluation results of HTTP/3 are inconclusive.